\documentstyle[12pt,epsfig]{article}
\def\lsim{\mathrel{\rlap{\lower3pt\hbox{\hskip0pt$\sim$}}
    \raise1pt\hbox{$<$}}}
\def\gsim{\mathrel{\rlap{\lower4pt\hbox{\hskip1pt$\sim$}}
    \raise1pt\hbox{$>$}}}
\begin{document}
\renewcommand{\theequation}{\thesection.\arabic{equation}}
\newcommand{\beq}{\begin{equation}}
\newcommand{\eeq}{\end{equation}}
\def\beqn{\begin{eqnarray}}
\def\eeqn{\end{eqnarray}}
\newcommand{\qt}{\tilde q}
\newcommand{\Tr}{{\rm Tr}\,}
\newcommand{\E}{{\cal E}}
\newcommand{\qtu}{\tilde q_{1}}
\newcommand{\qtd}{\tilde q_{2}}
\newcommand{\ntwo}{${\cal N}=2\;$}
\newcommand{\none}{${\cal N}=1\;$}
\newcommand{\ax}{\vert\xi\vert}
\newcommand{\xva}{\vert\vec{\xi}\vert}
\newcommand{\rt}{\tilde r}
\newcommand{\ar}{i\theta}
\newcommand{\as}{\alpha^{2}}
\newcommand{\gs}{g^{2}}
\newcommand{\fl}{\phi_{l}}
\newcommand{\fh}{\phi_{h}}
\newcommand{\pu}{\psi_{1}}
\newcommand{\pd}{\psi_{2}}
\newcommand{\vp}{\varphi}
\newcommand{\ve}{\varepsilon}
\newcommand{\pt}{\partial}
\newcommand{\pz}{\partial_{z}}


\begin{titlepage}
\renewcommand{\thefootnote}{\fnsymbol{footnote}}

\begin{flushright}
FTPI-MINN-03/35,
UMN-TH-2223/03,\\
ITEP-TH-95/03

\end{flushright}

\vfil

\begin{center}
\baselineskip20pt
{ \Large \bf Localization of Non-Abelian Gauge Fields on
Domain Walls at Weak Coupling \\
(D-brane Prototypes)}
\end{center}
\vfil

\begin{center}

\vspace{0.3cm}

{\large
{ \bf    M.~Shifman$^{a}$} and { \bf A.~Yung$^{a,b,c}$}
}

\vspace{0.3cm}

$^a${\it  William I. Fine Theoretical Physics Institute, University of Minnesota,
Minneapolis, MN 55455, USA}\\
$^b${\it Petersburg Nuclear Physics Institute, Gatchina, St. Petersburg
188300, Russia}\\
$^c${\it Institute of   Theoretical and Experimental Physics,
Moscow  117250, Russia}\\

\vfil

{\large\bf Abstract}
\end{center}
\vspace*{.25cm}

Building on our previous results (Ref.~\cite{Shifman:2002jm}),
we study  D-brane/string prototypes in weakly
coupled (3+1)-dimensional supersymmetric field theory engineered to support
(2+1)-dimensional domain walls, ``non-Abelian" strings
and various junctions.  Our main but not exclusive  task is the study
of localization  of  non-Abelian gauge fields on the walls.
The model we work with is  \ntwo QCD,
 with the gauge group SU(2)$\times$U(1)  and $N_f=4$ flavors
of fundamental hypermultiplets (referred to as quarks), perturbed by
the Fayet-Iliopoulos term of the U(1) factor.
In  the limit of large but almost
equal quark mass terms  a   set of vacua exists  in which this
theory is at weak coupling. We focus on these vacua (called the quark vacua).
We study elementary BPS domain walls interpolating between selected quark
vacua, as well as their bound state, a composite wall.
The composite wall is demonstrated to localize
a non-Abelian gauge field on its world sheet.
Next, we turn to the analysis of recently proposed
``non-Abelian" strings (flux tubes) which carry orientational moduli
corresponding to   rotations of the ``color-magnetic"
 flux direction  inside a global O(3). We find a
1/4-BPS solution for the  string ending on the composite domain wall.
The end point of this string is shown to play the role of a non-Abelian (dual)
charge in the effective world volume   theory of non-Abelian (2+1)-dimensional
vector fields  confined to the wall.

\vfil

\end{titlepage}

\newpage

\tableofcontents

\newpage

\section{Introduction}
\label{introduction}

String theory which emerged from dual hadronic models
in the late 1960's and 70's, elevated to the
``theory of everything" in the 1980's and 90's when it
experienced  an unprecedented expansion,  has
seemingly  entered a ``return-to-roots"
stage.  Results and techniques
of string/D-brane theory, being applied to non-Abelian field
theories (both, supersymmetric and non-supersymmetric),
have recently generated numerous predictions
for  gauge theories at strong coupling. If the latter are, in a sense, dual to
string/D-brane theory ---  as is generally believed to be the case ---
they must support  domain walls (of the D-brane type) \cite{P},
and we know, they do \cite{DS,Witten:1997ep}.
In addition, string/D-brane theory teaches us that a
fundamental string that starts on a confined quark, can end
on the domain wall \cite{P}.

In our previous paper \cite{Shifman:2002jm}
we embarked on the studies of
field-theoretic prototypes of D branes/strings. To this end
 we considered (2+1)-dimensional domain walls
in  (3+1)-dimensional \ntwo SQCD with  the SU(2)  gauge group
(and $N_f=2$ flavors of
fundamental hypermultiplets --- quarks),   perturbed by a small mass
term of the adjoint matter. In fact, our analysis reduced to that of
the effective low-energy \ntwo SQED with a (generalized) Fayet-Iliopoulos term.
 We found  1/2 BPS-saturated domain wall solution
interpolating between two quark vacua at  weak coupling.
The main finding was the {\em localization} of a U(1)  gauge field on this
domain wall.  We also demonstrated that  the Abrikosov-Nielsen-Olesen magnetic flux
tube  can end on the  wall.

The goal of the present work is the extension of the above results.
Now we want to consider composite walls,
analogs of a stack of D branes,  to see that
they localize non-Abelian gauge fields, say U(2).
Our second task --- as important to us as the first one ---
is the study of   non-Abelian flux tubes, and,
especially,  how they end on the walls.
In this way we continue the line of research
initiated in Refs.~\cite{Hanany:2003hp,Auzzi:2003fs,Tong:2003pz}.

The set up that will provide us with the appropriate tools is the same as in
the previous paper \cite{Shifman:2002jm}, namely \ntwo SQCD
analyzed by Seiberg and Witten \cite{SW1,SW2}.
Compared to Ref.~\cite{Shifman:2002jm} we will deal with a
somewhat different version, however.
We will start from the SU(3) theory with four
``quark" hypermultiplets ($N_f=4$) in the fundamental representation.
The SU(3) gauge group will be spontaneously broken down to
SU(2)$\times$U(1) at a large scale $m$ where
$m\sim m_A=m,\,\,\, A=1,2,3,4$,
and $m_1,\,\,  m_2, m_3$ and $m_4$ are the mass terms ascribed to the four quark flavors that are present in the model. Generically, all $m_A$'s are different, but we will choose a non-generic configuration.

Although SU(3) \ntwo SQCD provides a conceptual skeleton
for our set up, in essence its role is to stay behind the scene,
as a motivating factor.
Since the gauge SU(3) group will be broken
at the largest scale relevant to the model,
and the bulk of our present work refers to lower scales,
in practice our set up is based on SU(2)$\times$U(1) gauge model with
four quark hypermultiplets and unbroken \ntwo. The underlying
SU(3) \ntwo SQCD which one may or may not keep in mind in reading this paper,
will be referred to as the ``prototheory."

With four  flavors,  the  SU(2)  subsector  is conformal; therefore the problem we address can be fully analyzed in the weak coupling regime. In fact, two
of four quarks will be just
spectators while the other two will play a nontrivial role
in the solution. The role of the spectators is to ensure
the conformal regime (see below).

As was mentioned, we will deal with  the gauge symmetry
breaking pattern  of a  hierarchical type. First, at a large scale
$\sim m\gg\Lambda_{{\rm SU(3)}}$ the gauge group
SU(3) is broken down to a subgroup SU(2)$\times$U(1)
by the vacuum expectation values
(VEV's) of the adjoint scalars.\footnote{The generic pattern of the SU(3) gauge group breaking by the adjoint VEV's is
${\rm SU}(3)\to {\rm U}(1)\times {\rm U}(1)$. This case
essentially reduces to the problem which had been considered
previously \cite{DS,DV,Shifman:2002jm} ---
  localization of  the
Abelian gauge fields on the wall. Here we are
interested in localization of   the non-Abelian gauge fields.
 Therefore, we we will deal with a special regime in which
${\rm SU}(3)\to {\rm SU}(2)\times {\rm U}(1)$.
In Refs.~\cite{APS,CKM} (see also \cite{MY}) it was shown
that some of \none
vacua of SU$(N)$ \ntwo SQCD can preserve a non-Abelian subgroup.}
 (Here $\Lambda_{{\rm SU(3)}}$ is the dynamical scale
parameter of  SU(3)).
Second, at
a lower scale $\sim \sqrt{\mu\, m_A}$,
 in the presence of the  adjoint mass term $\mu\,{\rm Tr}\Phi^2$, the  light  squark fields acquire VEV's of  the   color-flavor diagonal form  (``color-flavor locking"),
\beq
  q_k^A     =   \delta_{A}^{k}   \,\sqrt{\mu \,  m}, \qquad  k,A =1,2.
\label{Qvev}
\eeq
In each vacuum to be
considered below, the squark fields of two (out of four) flavors
will be condensed, so that we can label each vacuum by a set of two
numbers, $(AB)$, showing that the flavors $A$ and $B$ are condensed.
For instance, we will speak of 12-vacuum, 13-vacuum and so on.

The basic idea of the gauge field
localization on the domain walls  is that the quark fields
(almost)  vanish inside the wall. Therefore, the gauge group SU(2)$\times$U(1),
being Higgsed in the vacua to the right and to the left of the wall,
 is restored inside the wall. Correspondingly, dual gauge bosons,
 being confined outside the wall are unconfined (or less confined),
 inside, thus leading to   localization \cite{DS}.

\begin{figure}
\epsfxsize=7cm
\centerline{\epsfbox{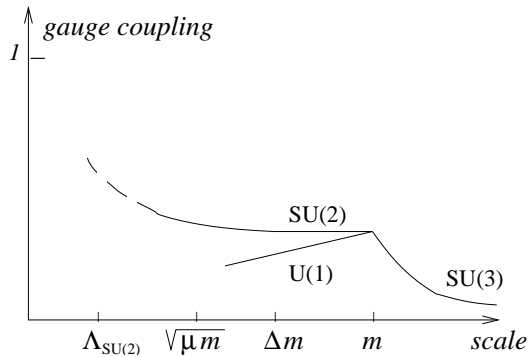}}
\caption{
The scale hierarchy: illustrating the fact that the gauge couplings
never become large in the problem at hand. According to
Eq.~(\ref{defxi}) $\mu\, m =\xi/6$.}
\label{hie}
\end{figure}

In fact, there is another scale in the problem which
plays an important role in  the aforementioned hierarchy.
In dealing with domain walls
we cannot consider the limit in which {\em all}   quark masses are
exactly equal. In this limit the pairs of appropriate
vacua coalesce, and  we
have no domain walls interpolating between them.
Therefore, we consider the   limit
of almost coinciding quark mass terms,
\beq
m_1=m_2,\quad m_3=m_4;\qquad
\Delta m \equiv m_1-m_3;\qquad |\Delta m | \ll m\,.
\label{massdiff}
\eeq
The resulting hierarchy
\begin{eqnarray}
&& |\Delta m|, \,\, m \gg \Lambda_{{\rm SU(3)}}\,,\nonumber\\[3mm]
&& \Lambda_{{\rm SU(2)}}\ll\sqrt{\mu m}\ll |\Delta m| \ll m
\label{rhierarch}
\end{eqnarray}
is exhibited in Fig.~\ref{hie},
together with the behavior of the corresponding gauge
couplings.
Here
\beq
m=\frac{m_1+m_3}{2}\,.
\label{emsred}
\eeq
Note that $\Lambda_{{\rm SU(2)}}$
is a would-be SU(2) dynamical scale.
It is of the order of
\beq
\Lambda_{{\rm SU(2)}}\sim |\Delta m|\exp\left(
-{4\pi^2}/{g^2_{{\rm SU(2)}}}\right)\,.
\label{dmscale}
\eeq
That's where
the SU(2) gauge coupling could have exploded. However, the problem under consideration is insensitive to this scale, as we will explain in detail in due course.

\begin{figure}
\epsfxsize=7cm
\centerline{\epsfbox{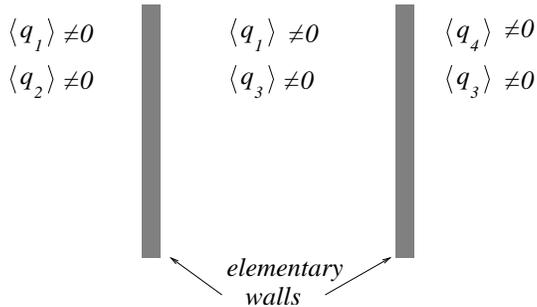}}
\caption{
Two elementary walls which comprise
a composite $12\to 34$ wall.}
\label{elemcompo}
\end{figure}

The theory at hand has domain walls of distinct types.
Assume that in the vacuum to the left of the
wall the squarks with the flavor indices 1 and 2 condense.
If in the vacuum to the right of the wall
the condensed squarks are 1 and 3, we will call such a wall
{\em elementary}. If, on the other hand,
the condensed squarks to the right of the wall are
3 and 4, this wall is obviously composite ---
it ``consists" of two elementary walls,
$12\to 13$ and $13\to 34$, see Fig.~\ref{elemcompo}.

The domain wall which localizes  SU(2)  gauge fields is {\em not}
elementary. It is a bound state of two elementary domain walls
 placed at one and the same position. This is in accordance
with the string/brane picture in which  SU(2)  gauge theory is
localized on the world volume of a stack of two coinciding   D-branes.
 If, however, the two  D-branes   are separated,
then  in string theory
the  SU(2)  gauge group is broken to U(1)$^2$, while  the masses of the
``charged" $W$ bosons are linear in the brane separations. We will recover
this picture in our field-theoretical set up.

The first stage of the spontaneous symmetry breaking,
${\rm SU}(3)\to {\rm SU}(2)\times {\rm U}(1)$,
is well studied in the literature, and presents no interest for our purposes.
Therefore, our dynamical analysis will start in essence
from the  ${\rm SU}(2)\times {\rm U}(1)$ model.
If one wishes, one can keep in mind that this latter
model is originally embedded in ${\rm SU}(3)$ \ntwo SQCD,
the ``prototheory," but this is not crucial.

Next, we turn to the analysis of recently proposed
non-Abelian strings (flux tubes) which carry orientational moduli
corresponding to  rotations of the ``color-magnetic"
 flux direction  inside a global SU(2)\cite{Auzzi:2003fs}
(similar results in three dimensions were obtained in
Ref.~\cite{Hanany:2003hp})\,.\footnote{A very fresh
publication \cite{ABEK}, which appeared after
the completion of the present paper, also examines
strings and their relation to monopoles.} We find a
1/4-BPS solutions for the non-Abelian string ending on the  composite
domain wall. The end point of the string is shown to play the role of a
non-Abelian charge in the effective world volume   theory of
non-Abelian (2+1)-dimensional gauge fields confined to the wall.

\vspace{1cm}

\section{Theoretical set-up: SU(2)$\times$ U(1) \ntwo  SQCD}
\setcounter{equation}{0}
\label{setup}

In this section we will describe the model we will work with
(a descendant of SU(3) Seiberg-Witten model with four
matter hypermultiplets) and
appropriate pairs of vacua which are connected by elementary
and composite walls.

\subsection{The model}
\label{model}

As was mentioned in Sect.~\ref{introduction},
the model we will deal with derives from
 \ntwo SQCD with the gauge group SU(3) and four  flavors of
the quark hypermultiplets. At a generic point on the Coulomb
branch of this theory, the gauge group is broken down to
U(1)$\times$U(1).  We will be interested, however,
in a particular subspace of the Coulomb branch, on which
the  gauge group is broken down to SU(2)$\times$U(1). We will
enforce\,\footnote{In certain vacua to be
considered in this paper the gauge group is further broken
to U(1)$\times$U(1) at a much lower scale $|\Delta m|$, see
Sect.~\ref{tvsales}.} this regime by a special choice of the
quark mass terms, see Eq.~(\ref{massdiff}).

The breaking SU(3)$\to$SU(2)$\times$U(1) occurs at the scale $m$
which is supposed to lie very high, $m\gg\Lambda_{\rm{SU(3)}}$.
Correspondingly, the masses of the gauge bosons from
SU(3)$/$SU(2)$\times$U(1) and their superpartners,
proportional to $m$, are very large, and so are the masses
of the third color component of the matter fields in the fundamental
representation.
We will be interested in phenomena at the scales $\ll m$.
Therefore, our starting point is in fact
the SU(2)$\times$U(1) model with four matter fields in the doublet
representation of SU(2),
as it emerges after the
SU(3)$\to$SU(2)$\times$U(1) breaking. These matter fields
are also coupled to the U(1) gauge field.

The field content
of SU(2)$\times$U(1) \ntwo SQCD with
four flavors  is as follows. The \ntwo vector multiplet
consists of the  U(1)
gauge fields $A_{\mu}$ and SU(2)  gauge field $A^a_{\mu}$,
(here $a=1,2,3$), their Weyl fermion superpartners
($\lambda^{1}_{\alpha}$,
 $\lambda^{2}_{\alpha}$) and
($\lambda^{1a}_{\alpha}$, $\lambda^{2a}_{\alpha}$), and
complex  scalar fields $a$, and $a^a$, the latter in the adjoint of
SU(2). The spinorial index of $\lambda$'s runs over
  $\alpha=1,2$.  In this sector the  global SU(2)$_R$ symmetry inherent to
  the model at hand manifests itself through rotations $\lambda^1 \leftrightarrow
  \lambda^2$.

The quark multiplets of  SU(2)$\times$U(1) theory consist
of   the complex scalar fields
$q^{kA}$ and $\tilde{q}_{Ak}$ (squarks) and
the  Weyl fermions $\psi^{kA}$ and
$\tilde{\psi}_{Ak}$,
 all in the fundamental representation of  SU(2)  gauge group.
Here $k=1,2$ is the color index
while $A$ is the flavor index, $A=1,2,3,4$.
Note that the scalars $q^{kA}$ and ${\bar{\tilde q}}^{\, kA}\equiv \overline{\tilde{q}_{Ak}}$
form a doublet under the action of the   global
 SU(2)$_R$ group.

The original SU(3) theory was perturbed by adding a small
mass term for the adjoint matter,  via the superpotential ${\cal W}=\mu{\rm Tr}\Phi^2$.
Generally speaking, this superpotential breaks
\ntwo down to ${\cal N}=1$.
The Coulomb branch shrinks to
a number of  {\em isolated} $\,$ \none vacua \cite{APS,CKM}.
In the limit of $\mu\to 0$ these vacua correspond to special
singular points on the Coulomb branch
in which  pair of monopoles/dyons or
quarks become massless.
The first three of these points (often referred to as the
Seiberg-Witten vacua) are always at
strong coupling. They correspond to \none vacua of   pure
SU(3) gauge theory.

The massless quark points ---
they present vacua of a distinct type,
to be referred to as the quark vacua --- may or may not be at weak
coupling depending on the values of the quark mass parameters  $m_A$.  If
$m_A\gg \Lambda_{{\rm SU}(3)}$, the quark vacua do lie at weak coupling.
Below we will be  interested only  in the  quark vacua
assuming that the condition $m_A\gg \Lambda_{{\rm SU}(3)}$ is met.

In the low-energy SU(2)$\times$U(1) theory, which is our starting point,
the perturbation ${\cal W}=\mu{\rm Tr}\Phi^2$ can be truncated, leading
to a crucial simplification. Indeed, since the ${\cal A}$
chiral superfield, the ${\cal N}=2$ superpartner of the U(1) gauge field,\footnote{ The superscript 2 in Eq.~(\ref{calasf}) is the global SU(2)$_R$ index of $\lambda$
rather than $\lambda$ squared. }
\beq
{\cal A} \equiv  a +\sqrt{2}\lambda^2\theta +F_a\,\theta^2\,,
\label{calasf}
\eeq
it not charged under the gauge group SU(2)$\times$U(1),
one can introduce the superpotential linear in ${\cal A}$,
\beq
{\cal W}_{{\cal A}} =-\frac1{\sqrt{2}}\, \xi\, {\cal A}\,.
\label{spla}
\eeq
It is rather obvious that ${\cal W}_{{\cal A}}$ is indeed a linear truncation
of ${\cal W}=\mu{\rm Tr}\Phi^2$.
A remarkable feature of the superpotential (\ref{spla})
is that it does {\em not} break \ntwo super\-symmetry
\cite{HSZ,VY}. Keeping higher order terms in $\mu{\rm Tr}\Phi^2$
would inevitably explicitly break \ntwo.
For our purposes it is crucial that the model we will deal with
is {\em exactly} \ntwo super\-symmetric.

The bosonic part of our SU(2)$\times$U(1)
 theory has  the form\,\footnote{Here and below we use a
formally  Euclidean notation, e.g.
$F_{\mu\nu}^2 = 2F_{0i}^2 + F_{ij}^2$,
$\, (\partial_\mu a)^2 = (\partial_0 a)^2 +(\partial_i a)^2$, etc.
This is appropriate since we are
going to study static (time-independent)
field configurations, and $A_0 =0$. Then the Euclidean action is
nothing but the energy functional. Furthermore, we
 define $\sigma^{\alpha\dot{\alpha}}=(1,-i\vec{\tau})$,
 $\bar{\sigma}_{\dot{\alpha}\alpha}=(1,i\vec{\tau})$. Lowing and raising of spinor indices
is performed by
virtue of the antisymmetric tensor defined as $\ve_{12}=\ve_{\dot{1}\dot{2}}=1$,
 $\ve^{12}=\ve^{\dot{1}\dot{2}}=-1$.
 The same raising and lowering convention applies to the flavor SU(2)
 indices $f$, $g$, etc. } \cite{Auzzi:2003fs}
\beqn
S&=&\int d^4x \left[\frac1{4g^2_2}
\left(F^{a}_{\mu\nu}\right)^2 +
\frac1{4g^2_1}\left(F_{\mu\nu}\right)^2
+
\frac1{g^2_2}\left|D_{\mu}a^a\right|^2 +\frac1{g^2_1}
\left|\partial_{\mu}a\right|^2 \right.
\nonumber\\[4mm]
&+&\left. \left|\nabla_{\mu}
q^{A}\right|^2 + \left|\nabla_{\mu} \bar{\tilde{q}}^{A}\right|^2
+V(q^A,\tilde{q}_A,a^a,a)\right]\,.
\label{qed}
\eeqn
Here $D_{\mu}$ is the covariant derivative in the adjoint representation
of  SU(2),
while
\beq
\nabla_\mu=\partial_\mu -\frac{i}{2}\; A_{\mu}
-i A^{a}_{\mu}\, \frac{\tau^a}{2},
\label{defnabla}
\eeq
where we suppress the color  SU(2)  indices, and $\tau^a$ are the
 SU(2) Pauli matrices. The coupling constants $g_1$ and $g_2$
correspond to the U(1)  and  SU(2)  sectors, respectively.
With our conventions the U(1) charges of the fundamental matter fields are $\pm
1/2$.

\vspace{1mm}

The potential $V(q^A,\tilde{q}_A,a^a,a)$ in the Lagrangian (\ref{qed})
is a sum of  $D$ and  $F$  terms,
\beqn
V(q^A,\tilde{q}_A,a^a,a) &=&
 \frac{g^2_2}{2}
\left( \frac{1}{g^2_2}\,  \varepsilon^{abc} \bar a^b a^c
 +
 \bar{q}_A\,\frac{\tau^a}{2} q^A -
\tilde{q}_A \frac{\tau^a}{2}\,\bar{\tilde{q}}^A\right)^2
\nonumber\\[3mm]
&+& \frac{g^2_1}{8}
\left(\bar{q}_A q^A - \tilde{q}_A \bar{\tilde{q}}^A\right)^2
\nonumber\\[3mm]
&+& \frac{g^2_2}{2}\left| \tilde{q}_A\tau^a q^A \right|^2+
\frac{g^2_1}{2}\left| \tilde{q}_A q^A - \xi \right|^2
\nonumber\\[3mm]
&+&\frac12\sum_{A=1}^4 \left\{ \left|(a+\sqrt{2}m_A +\tau^a a^a)q^A
\right|^2\right.
\nonumber\\[3mm]
&+&\left.
\left|(a+\sqrt{2}m_A +\tau^a a^a)\bar{\tilde{q}}_A
\right|^2 \right\}\,,
\label{pot}
\eeqn
where the sum over repeated flavor indices $A$ is implied,
and we introduced a {\em constant} $\xi$
related to $\mu$ as follows:
\beq
\xi = 6\, {\mu \, m}\,.
\label{defxi}
\eeq
The first and second lines represent   $D$   terms, the third line
the $F_{\cal A}$ terms,
while the fourth and the fifth  lines represent the squark $F$ terms.
As we know  \cite{HSZ,VY,EFMG,MY,Auzzi:2003fs},
this  theory supports BPS vortices.

Bearing in mind that we have
 four flavors we conclude that the  SU(2) coupling
does not run: SU(2) theory with $N_f=4$ is conformal.
Hence, $g^2_2$  is given
by its value at the scale $m$,
\beq
\label{g2}
\frac{8\pi^2}{ g^2_2}= 2\, \ln \frac{m}{\Lambda_{{\rm SU}(3)}}+\cdots.
\eeq
At large $m$ the  SU(2)  sector is indeed weakly coupled.

The  U(1)  coupling
undergoes an additional renormalization from scale $m$
down to the scale determined by the masses of light
states in the low-energy theory
(the latter are  of the order of $\sqrt{\mu m}\sim \sqrt\xi$, see Sect.~\ref{tvsales}).
At the scale $m$ the both couplings, $g_2^2$ and $g_1^2$
unify since at this scale they belong to SU(3). Note that
in passing from the SU(3) theory to SU(2)$\times$U(1)
we changed the normalization of the eighth
generator of SU(3) which became the generator of U(1),
see Eq.~(\ref{defnabla}). This change of normalization implies that
the unification condition takes the form
\beq
\frac{8\pi^2}{ g^2_1 (m)}= 3\, \frac{8\pi^2}{ g^2_2(m)}\,.
\label{normcond}
\eeq
The  one-loop coefficient of the
$\beta$ function
for  the U(1)  theory is $\beta_0=2\times 2\, n_e^2 N_f$
where the first factor of 2 reflects the difference in normalizations of
the SU$(N)$ versus U(1) generators,
the extra factor of two comes from the fact
that for each flavor we deal with matter doublets, and, finally,
the electric charge $n_e=1/2$, see Eq.~(\ref{defnabla}).
Thus, evolving $g^2_1$ from $m$ down to $\sqrt{\mu m}$
we get
\beq
\frac{8\pi^2}{ g^2_1 (\sqrt{\mu m})}=
6\, \ln \frac{m}{\Lambda_{{\rm SU}(3)}}
+2\, \ln \frac{m}{\mu }
 +\cdots ,
 \label{g1}
\eeq
Clearly,
this coupling is even smaller than that of the   SU(2)  sector.

To make readers'  journey through this work easier we display in
Table \ref{table1} the field content of the model.

\vspace{3mm}

\begin{table}
\begin{center}
\begin{tabular}{|c|cc|cc|cc|}
\hline
$\frac{\mbox{ SU(2)$_C$ repr.} \to  }{\mbox{  Spin} \downarrow} $ \rule{0cm}{0.7cm} &  singlet &   &  fundamental& /anti     & adjoint &  \\[3mm]
\hline\hline
\rule{0cm}{0.7cm}0  &  &$a$& $q^A$ &$\tilde q_A$  & &$a^a$
\\[2mm]
\hline
\rule{0cm}{0.7cm}1/2 &$\lambda^1$ &$\lambda^2 $& $\psi^A$&
$\tilde\psi_A$& $\lambda^{1a}$& $\lambda^{2a}$\\[2mm]
\hline
\rule{0cm}{0.7cm}1 &$A_\mu$ &&  &
 & $A_\mu^{a}$& \\[2mm]
\hline
\end{tabular}
\end{center}
\caption{Field content of the model under consideration.}
\label{table1}
\end{table}

\begin{table}
\begin{center}
\begin{tabular}{|c|c |}
\hline
${\cal N}=2$ SUSY  &  unbroken
  \\[3mm]
\hline
SU(2)$_R$  & unbroken
\\[2mm]
\hline
\{U(1)$\times$ SU(2)\}$_G\times$ SU(2)$_{f_{12}}\times$ SU(2)$_{f_{34}}$
$\times$ U(1) &
$\to \left\{\begin{array}{lll}
{\rm U}(1)_{\rm diag}\times{\rm SU}(2)_{\rm diag}
\times {\rm SU(2)}_{f_{34}}\,,\\ [1mm]
\quad 12\,{\rm vacuum;}\\[3mm]
{\rm U(1)}_{\rm diag}\,, \quad 13\, {\rm vacuum}
\end{array}\right.$
\\
\hline
\end{tabular}
\end{center}
\caption{Pattern of the symmetry breaking.}
\label{table2}
\end{table}

It is also instructive to summarize the symmetries of the model and
patterns of their breaking, see Table~\ref{table2}. Besides the gauge symmetries, of importance are the global symmetries of the model.
Our ``proto-SU(3)-model" (mentioned in passing)
had SU(3)$_c$ broken  down to
SU(2)$\times$U(1).  This breaking occurs at at the scale $m$.
The resulting superpotential
\begin{eqnarray}
{\cal W}& =& \frac1{\sqrt{2}}\sum_{A=1}^4\left( \tilde q_A  {\cal A}q^A  +
\tilde q_A  {\cal A}^a\,\tau^a q^A\right)
\nonumber\\[3mm]
&+& m_1\sum_{A=1,2}\tilde q_Aq^A    +
m_3\sum_{A=3,4}\tilde q_A q^A   \nonumber\\[3mm]
&-&\frac1{\sqrt{2}}\, \xi\, {\cal A}\,,
\label{sprptntl}
\end{eqnarray}
has, in addition, a large global  SU(2)$_{f_{12}}\times$ SU(2)$_{f_{34}}$
$\times$U(1) flavor symmetry.
At the scale $\Delta m$ the color
SU(2) may or may not be broken.
In the 13, 14, 23, and 24 vacua it is broken by
the vacuum expectation value of
$a^3$ down to U(1), paving the way to monopoles with typical sizes
$\sim (\Delta m)^{-1}$.
In the 12 and 34 vacua $\langle a^3\rangle $ does not develop,
and we can descend further, down the the scale $\xi$.
At this scale all gauge symmetries, in all six vacua
under consideration,  are fully Higgsed. The
Abrikosov-Nielsen-Olesen (ANO, Ref.~\cite{ANO})
 strings  are supported. The transverse size of these strings, $\sim \xi^{-1/2}$, is much larger than $  (\Delta m)^{-1}$. In fact, we will deal with two
distinct types of strings which generalize
their more primitive ANO counterparts. They
correspond (in the quasiclassical limit) to distinct types of winding, see Sect.~\ref{naftin2} for further details.

\subsection{The vacuum structure and excitation spectrum}
\label{tvsales}

This section briefly outlines  the vacuum structure and
the excitation mass spectrum of our basic
SU(2)$\times$U(1)  model  (for
further details, including those referring to the full SU(3) theory, see Refs.~\cite{APS,CKM,MY,Auzzi:2003fs}). First, we will
examine  relevant vacua.

The  vacua of the   theory (\ref{qed}) are determined
by the zeros of the potential (\ref{pot}).
We will assume   the conditions (\ref{rhierarch}) to be met. Then,
besides three strong-coupling vacua which exist  in   pure  SU(3)
\ntwo Yang-Mills
theory, we have eight vacua in which one quark flavor
is condensed, and six  vacua  in which two quark flavors  develop  non-vanishing
VEV's. For our problem --- domain walls and flux tubes
at weak coupling --- we  will choose these latter six vacua.
They are 12-, 34-, 13-, 14-, 23- and  24-vacua. In the first two SU(2) gauge
 symmetry is unbroken by adjoint scalars while in the last four vacua it is
broken by them at the scale $\Delta m$.

\begin{figure}
\epsfxsize=7cm
\centerline{\epsfbox{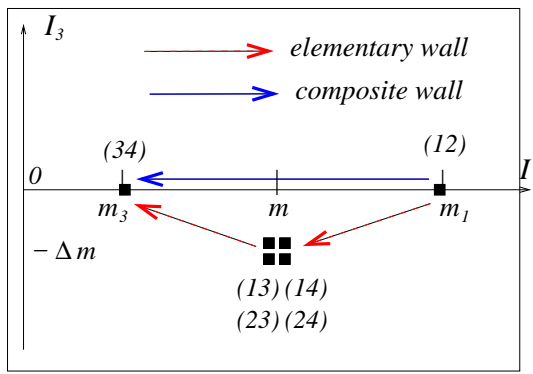}}
\caption{
The vacuum structure on the $(I\,,\,\, I_3)$ plane.}
\label{vac}
\end{figure}

Each of the above six vacua (AB) is labeled by a triplet of
gauge-invariant order parameters $I_1$, $I$ and $I_3$ defined as
\beqn
&&\langle \tilde q_A \,q^B\rangle \equiv
I_1\,\delta_A^B \,,
\nonumber\\[3mm]
&&\langle \tilde q_A \, a \,  q^B\rangle
\equiv  - \sqrt{2}\,
I \times I_1\,\delta_A^B \,,
\nonumber\\[3mm]
&&\langle \tilde q_A \, a^a\tau_a\, q^B\rangle
\equiv \frac{1}{\sqrt 2}\,
I_3  \times I_1\,(\tau_3)_A^B \,,
\label{odpar}
\eeqn
where summation over the SU(2) color indices is implied. The corresponding
vacuum structure
is exhibited in Fig.~\ref{vac}.

\subsubsection{SU(2)  symmetric vacua}
\label{su2sym}

Let us first  consider the  12 (or 34) vacuum.
The adjoint fields develop the following VEV's:
\beq
\langle a^3\rangle =0, \qquad \langle a\rangle =- \sqrt{2} \,
m_1 =- \sqrt{2} \, m
+O(\Delta m ) \,,
\label{avev}
\eeq
where $m$ is defined in Eq.~(\ref{emsred}).
If the  values of the mass parameters  $m_{1,3}$ and  $\mu$
are real, we can exploit  the freedom of
rotations in SU(2) and U(1)  to
make the quark VEV's real too. Then
in the case at hand they take the color-flavor locked form
\beqn
\langle q^{kA}\rangle &=&\langle \bar{\tilde{q}}^{kA}\rangle =\sqrt{
\frac{\xi}{2}}\, \left(
\begin{array}{cc}
1 & 0 \\
0 & 1\\
\end{array}
\right),
\nonumber\\[4mm]
k&=&1,2,\qquad A=1,2\,.
\label{qvev}
\eeqn
This particular form of the squark condensates is dictated by the
 third line in Eq.~(\ref{pot}). Note that  the squark fields stabilize
 at non-vanishing values
entirely due to the U(1) factor --- the second term in the third line.

The gauge invariants corresponding to the vacuum (\ref{qvev})
are
\beq
I_1 =\frac{\xi}{2}\,,\quad I = m_1\,,\quad I_3 = 0\,.
\label{gi12}
\eeq
At first site it might seem that, say,  the field configuration
\beqn
\langle q^{kA}\rangle &=&\langle \bar{\tilde{q}}^{kA}\rangle =\sqrt{
\frac{\xi}{2}}\, \left(
\begin{array}{cc}
0 & 1 \\
1 & 0\\
\end{array}
\right),
\nonumber\\[4mm]
k&=&1,2,\qquad A=1,2\,,
\label{qvevp}
\eeqn
which also provides a (12)-vacuum solution,
presents another vacuum. This is obviously not the case, since it is nothing but
a gauge copy of Eq.~(\ref{qvev}). The gauge invariants obtained from
Eq.~(\ref{qvevp}) are the same as in Eq.~(\ref{gi12}).

Let us move on to  the issue of the excitation spectrum  in this vacuum.
The mass matrix for the gauge fields $(A^{a}_{\mu},
A_{\mu})$ can be read off from the quark kinetic terms  in Eq.~(\ref{qed})
and has the form
\beq
{\cal M}_{V}^2 = \xi\, \left(
\begin{array}{cc}
  g_2^2 & 0  \\
0 & g_1^2  \\
  \end{array}\right)
  \label{r2phmatp}
\eeq
Thus, all three SU(2) gauge bosons become massive, with
one and the same mass
\beq
M_{1,2,3} =g_2\,\sqrt\xi\,.
\label{m123}
\eeq
The equality of the masses is no accident. The point is
that our model actually has the symmetry
SU(2)$_{c}\times$SU(2)$_{f_{1,2}}\times$SU(2)$_{f_{3,4}}$
where the first flavor SU(2) corresponds to rotations of the $A=1,2$
flavors, while the second flavor SU(2) to rotations of the $A=3,4$ flavors.
The pattern of the spontaneous breaking is such
that the diagonal SU(2) from the product
SU(2)$_{c}\times$SU(2)$_{f_{1,2}}$
remains an unbroken global  SU(2) symmetry of the theory.  Sure enough, SU(2)$_{f_{3,4}}$ is also unbroken, since the $A=3,4$ flavors play
a passive role of spectators in the 12-vacuum.

The mass of the U(1) gauge boson is
\beq
M_{{\rm U}(1)} =g_1\,\sqrt\xi\,.
\label{mu1}
\eeq
From the mass scale $m$ down to the  scale $|\Delta m |$
the gauge coupling $g_2^2$ does not run because of conformality.
Below $|\Delta m|$, the conformality is broken:
two quark flavors out of four have mass of the order of
$\Delta m$ and, hence,
decouple. Therefore, $g_2^2$ runs, generating a
dynamical mass scale (\ref{dmscale}).
At the mass scale $M_{1,2,3} =g_2\,\sqrt\xi$
this last running gets frozen.  Since $M_{1,2,3} \gg
\Lambda_{{\rm SU}(2)}$, by assumption,
the running of $g_2^2$ in the interval from $|\Delta m|$
down to $\sqrt\xi$ can be neglected.
Therefore,   we can treat $g_2^2$ in the above relations as a scale
independent constant (coinciding with the
gauge constant normalized at $m$). The mass spectrum of the
adjoint scalar excitations is the same as for the
gauge bosons. This is enforced by \ntwo.

What is the mass spectrum of the quark (squark) excitations?
These fields are color doublets. To ease the notation
it will be convenient (sometimes) to use subscripts $r$ and $b$
(red and blue) for the color indices of $q$ and $\tilde q$.
It is rather obvious that $q^{(A=1)}_r$,
$\, q^{(A=2)}_b$, $\, \tilde q_{(A=1)}^{\, r}$ and
$\, \tilde q_{(A=2)}^{\, b}$
are ``eaten up" in the Higgs mechanism.
The remaining four superfields, $q^{(A=2)}_r$, $\, q^{(A=1)}_b$, $\, \tilde q_{(A=2)}^{\, r}$ and $\, \tilde q_{(A=1)}^{\, b}$ split into two groups ---
a singlet under the residual global SU(2) with the mass
(\ref{mu1}), and a triplet under the residual global SU(2) with the mass
(\ref{m123}). Altogether we have 1+3 =4 long massive \ntwo
supermultiplets with mass squared proportional to $\xi$.

As for the spectator quark flavors
(those that do not condense in the given vacuum),
the quarks/squarks of the third and fourth flavors
are much heavier. They
have masses   $\sim |\Delta m|$,
as is clear from Eq.~(\ref{pot}).
Assuming the limit (\ref{massdiff}), we
include the
spectator quarks/squarks
 in the low-energy theory (\ref{qed}).
In particular, each spectator quark flavor with the mass term $m_3$
--- remember, we have two of those ---
produces two \ntwo multiplets.
The first one, with the mass  $|m_3-m_1|= |\Delta m|$, is formed from the
$r$-components of the spectator quark, while the second one,
with the same mass,  is formed from its $b$-components.
Thus, each  supermultiplet contains four bosonic and four fermionic (real)
degrees of freedom (short \ntwo supermultiplet).  Altogether we have four   such  supermultiplets --- two doublets of the
global SU(2). There are no massless excitations in this vacuum.

The 34-vacuum is similar. The only difference is that
the value of the gauge invariant $I$ in the 34-vacuum is $I=m_3$.

\subsubsection{SU(2)  non-symmetric vacua}
\label{su2nsym}

As an example of such vacuum we will consider the 13-vacuum.
Now, in contradistinction
with  the previous case, both adjoint fields, $\langle a^3\rangle$ and $\langle a\rangle$, develop vacuum expectation values, so that
Eq.~(\ref{avev})
must be replaced by
\beqn
\langle a^3\rangle &=& -\frac{m_1-m_3}{\sqrt 2}\equiv
-\frac{\Delta m }{\sqrt 2}\,,
\nonumber\\[3mm]
\langle a \rangle &=& -\frac{m_1+ m_3}{\sqrt 2}\equiv  -    {\sqrt 2}\, m \,.
\label{avevns}
\eeqn
The above vacuum values of the adjoint scalars
follow from examination of the last two lines in Eq.~(\ref{pot}).
The squark fields in the vacuum are similar to those
in Eq.~(\ref{qvev}), with the replacement of the second flavor by the third one,
namely,
\beqn
\langle q^{kA}\rangle &=&\langle \bar{\tilde{q}}^{kA}\rangle =\sqrt{
\frac{\xi}{2}}\, \left(
\begin{array}{cc}
1 & 0 \\
0 & 1\\
\end{array}
\right),
\nonumber\\[4mm]
k&=&1,2,\qquad A=1,3\,,
\label{qvev13}
\eeqn
up to gauge copies. The gauge invariant order parameters are
\beq
I_1 =\frac{\xi}{2}\,,\quad I = m \,,\quad I_3 = - \Delta m  \,,
\label{gi13}
\eeq
see Fig.~\ref{vac}.

Next, let us examine the excitation spectrum in this vacuum.
In this vacuum the gauge group of our model is fully Higgsed too ---
all four gauge bosons acquire masses. No ``custodial" global
SU(2) survives, however. Correspondingly,
the masses of the gauge bosons $A_\mu^{1\pm i2}$ on the one hand, and
$A_\mu^3$ on the other, split.

More concretely,
\beq
M(A_\mu^{1\pm i2}) = M(a^{1\pm i2}) = \Delta m  \gg \xi\,,
\label{mcharge}
\eeq
while the masses  of $A_\mu$
and   $A_\mu^3$ (and the same for $a$
and   $a^3$) are given by the same values as in the SU(2)-symmetric vacua,
\beqn
M_{{\rm U}(1)} & =& g_1\,\sqrt\xi\, ,
\nonumber\\[3mm]
M_{3} & =& g_2\,\sqrt\xi\,.
\label{r2mass}
\eeqn

The mass matrix for the lightest quarks has the size
$8\times 8$, including four
(real) components of the $q_{r}^1$ quark and four components
of the $q_{b}^3$ quark.
It has two vanishing eigenvalues
associated with   two states ``eaten'' by the Higgs
mechanism for two U(1)  gauge factors,  and two non-zero eigenvalues
coinciding  with   masses (\ref{r2mass}). Each of these non-zero eigenvalues
corresponds to three quark eigenvectors.
Altogether we have two long \ntwo multiplets with masses
 (\ref{r2mass}), each one containing eight bosonic and eight fermionic states.

Let us remember  that, in the limit $m_1=m_2$ and $m_3=m_4$,
the 13-vacuum coalesce
with three others, namely, the 14-, 23- and 24-vacua, see Fig.~\ref{vac}.
This means that we have massless multiplets in these vacua. In fact,
the common position of these vacua on the Coulomb branch is the  root of
a Higgs branch.
This  Higgs branch has dimension eight, cf. \cite{APS,MY}. To see
this observe, that in the $(A,B)$ vacuum with $A=1,2$ and
$B=3,4$ we have 16
real quark scalar variables ($q_{r}^1$, $q_{r}^2$, $q_{b}^3$ and $q_{b}^4$)
 subject to two $D$-term
conditions and four $F$-term conditions. Also we have to subtract
two U(1) phases. Overall we have $16-6-2=8$ which gives us the
dimension of the Higgs branch. This dimension should be a multiple of four
since the Higgs branches are hyper-K\"{a}hler manifolds \cite{SW2}.

\section{The
\boldmath{${\cal N}=2$}
 central charges   relevant
to the problem}
\label{ccn2}
\setcounter{equation}{0}

The model under consideration supports, in various limits,
all three classes of topological defects that are under  scrutiny
in the current literature:
domain walls, strings and monopoles.
Below we will explore    BPS-saturated defects,
with a special emphasis on various junctions. The domain walls
and strings are 1/2 BPS, the wall-string junctions and the string-string
junctions are 1/4 BPS.

It is instructive to begin from the discussion of corresponding central
charges in ${\cal N}=2$ superalgebra. While a part
of the material below is a mini-review,
in the analysis of the monopole central charge we will add a bifermion term
which was routinely
omitted previously. It was omitted for a good reason, though:
for a free monopole the contribution of this bifermion term
vanishes. It is   crucial, however, for the
confined monopoles to which we will turn below.

\subsection{$ (1,0)$ and $(0,1)$ central charges}
\label{01cc}

These central charges are saturated by domain walls.
They appear in the anticommutators
$\{Q_\alpha^f\, Q_\beta^g\}$ (remember, $f,g =1,2$ are SU(2)$_R$ indices).
Since $\{Q_\alpha^f\, Q_\beta^g\} \sim \int d\,  \Sigma_{\alpha\beta}$
where $d\, \Sigma_{\alpha\beta}$ is the element of the area of the domain wall in question ($d\Sigma_{\alpha\beta} = d\Sigma_{\beta\alpha} $),
the (1/2,1/2) central charges must be symmetric with respect to the
interchange $f\leftrightarrow g$. More precisely,
\beq
\{Q_\alpha^f\, Q_\beta^g\} = -4\,\Sigma_{\alpha\beta}\, \bar{\cal Z}^{fg}\,,
\eeq
where
\beq
\Sigma_{\alpha\beta} =-\frac{1}{2}\,\int_{\rm wall}\, dx_{[\mu} dx_{\nu]}
(\sigma^\mu)_{\alpha\dot\alpha}\,(\bar\sigma^\nu)^{\dot\alpha}_\beta
\sim {\rm Wall \,\,\, Area}\,,
\eeq
while the central charge $\bar{\cal Z}^{fg}$ ought to be an SU(2)$_R$ vector.

In the string-theory context the $ (1,0)$ and $(0,1)$ central charges were first
discussed in Ref.~\cite{deAzcarraga:gm}.
The discovery of the corresponding field-theoretic anomaly in
supersymmetric Yang-Mills
(SYM) theories \cite{DS} paved the way to multiple explorations and uses of the
$ (1,0)$ and $(0,1)$ central charges in field theory (e.g. \cite{KSS,Chibisov:1997rc,Ritz:2002fm} to name just a few).

Two observations severely constrain the form of the central charge $\bar{\cal Z}^{fg}$: first, it must be (anti)holomorphic in fields;
second, it must be a SU(2)$_R$ vector.
As a result, the most general form compatible with the above observations
is\,\footnote{Derivation of Eq.~(\ref{stenkaz})
also exploits the specific feature of \ntwo theories
that each given superfield enters in the superpotential linearly,
see Eq.~(\ref{sprptntl}). This implies, in particular, that
in any given vacuum (at the classical level)
${\cal W}_2 + {\cal W}_3 \equiv 0$ where ${\cal W}_{2,3}$ are the quadratic and cubic parts of the superpotential.}
\beq
{\cal Z}^{fg} = \Delta \left(-  \frac{4}{\sqrt 2}\,  \xi^{fg}\,  {\cal A}
+   \frac{c_1}{16\pi^2}\,   \lambda^f \lambda^g
+ \frac{c_2}{16\pi^2}\,   \lambda^{af} \lambda^{ag}\right)\,,
\label{stenkaz}
\eeq
where $\Delta$ means the difference of the expectation
values of the operator in parentheses in two vacua
between which the wall in question interpolates. Furthermore,
the parameter $\bar\xi^{fg}$ is an SU(2)$_R$
matrix (related to a real vector $\vec\xi\, $) introduced in Ref.~\cite{VY}.
In the model under consideration
$ \xi^{fg} = ( \xi /2 )\, {\rm diag}\, \{1, -1\}$, see the remark after
Eq.~(\ref{1212cc}).

The last two terms in Eq.~(\ref{stenkaz}), containing numerical
coefficients $c_{1,2}$, present a quantum anomaly, a generalization of
that of Ref.~\cite{DS}.
The coefficients $c_{1,2}$ are readily calculable in terms of the
Casimir operators of the gauge group of the model under consideration;
they also depend on the matter content.
We will not dwell on them here because, given our
hierarchy of parameters
(\ref{rhierarch}), the anomalous terms
in $ {\cal Z}^{fg}$ will play no role.
A rather straightforward algebra
(in conjunction with known results) yields us the coefficient
in front of $ \xi^{fg}\,  {\cal A}$ quoted in Eq.~(\ref{stenkaz}).
To this end we combine Eq.~(\ref{sprptntl}) above
with Eqs. (3.19) and (3.20) from Ref.~\cite{Shifman:1999mv}.
In our normalization the BPS wall tension reduces\,\footnote{For generic matrices $ {\cal Z}^{fg}$ it is the eigenvalue of ${\cal Z}^{fg}$ that counts.}  e.g. to
$T_{\rm w} =| {\cal Z}^{11} |$.

\subsection{(1/2, 1/2) central charge}
\label{12cc}

This central charge is saturated by strings
(flux tubes).\footnote{It is also instrumental in the issue of
BPS-saturated vortices  \cite{Chibisov:1997rc} and wall junctions \cite{GS,Gabadadze:1999pp}. One can trace this
line of reasoning to Ref.~\cite{at}.}
It appears in the anticommutator
$\{Q_\alpha^f\, \bar Q_{\dot\beta\, g}\}$. This central charge
is not holomorphic,
and has no particular symmetry with respect to
permutations of the SU(2)$_R$ indices $f$ and $g$.

It is well-known that the (1/2, 1/2) central charge exists also in
  \none  supersym\-metric QED (SQED) with the Fayet-Iliopoulos term,
see Ref. \cite{HS} and
especially Ref. \cite{DDT}, specifically devoted
to this
issue. In Ref. \cite{DDT} it is shown, in particular,
that if the spontaneous breaking of U(1) is
due to the Fayet-Iliopoulos term  \cite{FI},
 then the corresponding ANO string is saturated in \none,
and the string  tension is given
by the value of the central charge. In Ref.~\cite{GS}
it was proven that at weak coupling
this is the {\em only} mechanism leading to BPS strings in \none theories,
which are, thus, by necessity, the ANO strings.\footnote{Note, however, that
the (1/2,1/2) central charge is missing in the
general analysis of Ref.~\cite{FP}.}

Our emphasis will be on ``non-Abelian,"
rather than ANO strings (the meaning of ``non-Abelian-ness" is explained in Sect.~\ref{naftin2}). It is only the extended,  \ntwo, supersymmetry
that can make them BPS-saturated.
Following Ref.~\cite{VY}, the (1/2,1/2)
anticommutator  in the \ntwo model at hand can be written as follows:
\beq
\{Q_\alpha^f\, \bar Q_{\dot\beta\, g}\} =
2\, \delta^f_g    \,     (\sigma^{\mu})_{\alpha\dot\beta}\, P_{\mu}
+4i \,     (\sigma^{\mu})_{\alpha\dot\beta}\,     \xi^f_g\,   \int d^3 x \, \frac12
\,  \varepsilon_{0\mu\delta\gamma}\,   F_{\delta\gamma},
\label{1212cc}
\eeq
where $P_{\mu}$ is energy-momentum operator while\,\footnote{
Note that the definition of $\xi^f_g$ in Ref.~\cite{VY}
differs by a factor of 2.}
$$
\xi^f_g=(\tau^m/2)^f_g\, \xi^m\,.
$$
Moreover, the  vector $\xi^m$ is a SU(2)$_R$ triplet
of generalized  Fayet-Iliopoulos
parameters (in our model only $\xi^1\equiv \xi$ is non-zero).

The second term in Eq.~(\ref{1212cc}) is the (1/2, 1/2) central charge.
It is worth emphasizing that it is only the
U(1) field $F_{\delta\gamma}$ that enters; the SU(2) gauge field
does not contribute to this central charge for rather evident reasons.
The central charge is obviously proportional to $L$,
the length of the string, times  the
magnetic flux of the string $\int d^2 x \vec B$ directed along the
string axis. With the normalizations accepted throughout this paper
one can write for the BPS string tension
\beq
T_{\rm s} = \left|\int d^2 x \vec B\right| \, |\, \vec \xi \, | = 2\pi \xi \,,
\eeq
where the last equality refers to the elementary
strings. For the ANO string the flux is twice larger, so that
$T_{\rm ANO}=4 \pi \xi$, see Sect.~\ref{naftin2}.

\subsection{The Lorentz-scalar  central charge}
\label{lsmcc}

As well-known \cite{Haag:1974qh}, this central charge is
possible only because of the extended nature of supersymmetry, \ntwo.
It appears in the anticommutator
$\{Q_\alpha^f\,  Q_{\ \beta}^g\}$ and has the structure
\beq
\{Q_\alpha^f\,  Q_{\ \beta}^g\} =\varepsilon_{\alpha\beta}\,
\varepsilon^{fg}\,2\,  Z\,,
\eeq
where $Z$ is an SU(2)$_R$ singlet while the factor of 2 on
the right-hand side is a traditional normalization. It is most convenient to write $Z$
in terms of  the topological charge, an integral over the topological density,
\beq
Z = \int \, d^3 x \, \zeta^0 (x)\,.
\eeq
In the model at hand
\beqn
\zeta^\mu &=& \frac12\varepsilon^{\mu\nu\rho\sigma}
\,\partial_\nu\left(
\frac{i}{g_2^2}\,   a^a F^a_{\rho\sigma} + \frac{i}{g_1^2}\,
 a F_{\rho\sigma}\right.
\nonumber\\[3mm]
&+& \left. \frac{c}{4 \pi^2}\, \lambda^f_\alpha\, s^{\alpha\beta} \,
\lambda^g_\beta \, \varepsilon_{fg} +\frac{2\, c\, g_2^2 }{4\pi^2}\,
\psi^{A}_{\alpha}\, s^{\alpha\beta} \, \tilde\psi_{A\, \beta} \right)\,,
\label{lscce}
\eeqn
where $s_{\alpha\beta}$  is a symmetric matrix corresponding to the
Lorentz representation (1,0) in the spinorial notation,
\beq
 s_{\alpha\beta} = (\sigma^{[\rho})_{\alpha\dot\alpha}\,
(\bar\sigma^{\sigma]})^{\dot\alpha}_\beta\,,
\eeq
while the  square brackets in the superscripts
 denote antisymmetrization with respect
to $\rho$ and $\sigma$. Moreover, $c$ is a numerical coefficient.

Two comments are in order here.  First,
the first two terms in Eq.~(\ref{lscce}) present the  conventional ``monopole" central charge which is routinely discussed in numerous reviews. It emerges
from the canonic (anti)commutators at the tree level.
In generic models it is in fact the (1,0) gauge field strength tensor
which appears in these classical terms in the first line in Eq.~(\ref{lscce}),
i.e. the (anti)selfdual combination. In the model at hand only the magnetic field survives in the expression for the central charge;
therefore, we dropped the electric component.

The last two (bifermion) terms in Eq.~(\ref{lscce}) are due to an
anomaly, which  is, in a sense,  an \ntwo counter-partner to that of Ref.~\cite{DS}. They will be discussed in more detail in the
accompanying paper \cite{SYfut} where the value of the
coefficient $c$ will be determined. They were unknown
previously playing no role in the routine monopole analysis.
They do play a crucial role,  however, for the Higgs phase monopoles
(confined monopoles), to be discussed in brief in Sect.~\ref{swjunc}.
In fact, this anomaly must match the recently obtained anomalous central charge \cite{Losev:2003gs} in two-dimensional O(3) sigma model.
More on that will be said in Sect.~\ref{swjunc} and Ref.~\cite{SYfut}.

\section{Domain walls}
\label{domwa}
\setcounter{equation}{0}

In this section we study BPS domain walls between various vacua
described in Sects.~\ref{su2sym} and \ref{su2nsym}.
First, we derive the first order equations for the
BPS walls and, second, find and analyze their
solutions. Our  final goal is to work out the solution for the composite wall
$12\to 34$ on  which we will eventually get   localized   non-Abelian gauge fields.

\subsection{First-order equations for elementary and composite walls}
\label{foe}

Let us note that the structure of the vacuum
condensates in all six  vacua considered in Sects.~\ref{su2sym} and \ref{su2nsym}
suggests that we can search for the domain wall solutions using the
{\em ansatz}
\beq
q^{kA}=\bar{\qt}^{kA}\equiv
\frac{1}{\sqrt{2}}\,\vp^{kA}\, ,
\label{qqt}
\eeq
where we introduce a new complex
field  $\vp^{kA}$, $k=r,b$.
Note that the above {\em ansatz} violates holomorphy in the space of fields
inherent to $F$ terms: superpotentials and certain other expressions
derivable from them. This is why some expressions presented below
which should be holomorphic on general grounds, do not look
holomorphic on the {\em ansatz} (\ref{qqt}).

Within this {\em ansatz} the effective action (\ref{qed}) becomes
\beqn
S &=& \int {\rm d}^4x\left\{\frac1{4g_2^2}
\left(F^{a}_{\mu\nu}\right)^{2}
+ \frac1{4g_1^2}\left(F_{\mu\nu}\right)^{2}
 + \frac1{g^2_2}\left|D_{\mu}a^a\right|^2 +\frac1{g^2_1}
\left|\partial_{\mu}a\right|^2
\right.
\nonumber\\[3mm]
& +&  \left. |\nabla_\mu \varphi^A|^2
+ \frac{g^2_2}{8} \left(
\bar{\varphi}_A\tau^a \vp^A \right)^2
 + \frac{g^2_1}{8}\left(
|\varphi^A|^2 -2\xi\right)^2\right.
\nonumber\\[3mm]
& +&
\left.\frac12\left|(a^a\tau^a+a
+\sqrt{2}m_A)\vp^A \right|^2
 \right\}\,,
\label{redqed}
\eeqn
where we use the same notation as in Eq.~(\ref{defnabla}).

For the time being let us  drop
the gauge field in Eq.~(\ref{redqed}).
It is irrelevant for the ``standard" domain wall.
If we  assume that all fields depend only on the
coordinate $z\equiv x_3$,
the Bogomolny completion\,\footnote{The Bogomolny completion
is routinely used in such problems after its introduction in Ref.~\cite{B}
and   the subsequent identification of the  central charges
of various  superalgebras with  topological charges ~\cite{OW}.}
 of the wall energy functional can be written as
\begin{eqnarray}
T_{\rm w} &=& \int dz \left\{
\left|\pt_z \vp^A\pm
\frac1{\sqrt{2}}(a^a\tau^a+a+
\sqrt{2}m_A)\vp^A\right|^2
\right.
\nonumber\\[3mm]
&+&
\left|\frac1{g_2}\pt_z a^a \pm \frac{g_2}{2\sqrt{2}}
(\bar{\varphi}_A\tau^a \vp^A)\right|^2
\nonumber\\[3mm]
&+&
\left. \left|\frac1{g_1}\pz a \pm \frac{g_1}{2\sqrt{2}}
(|\vp^A|^2-2\xi)\right|^2
\,\, \pm  \sqrt{2}\xi\pz a  \right\}.
\label{bog}
\end{eqnarray}
In the above expression we have omitted another full-derivative
boundary term proportional to
$(\partial /\partial z )\sum Q (\partial W /\partial Q)$
where the sum runs over all superfields. Since in all vacua
$(\partial W /\partial Q) =0$, this term produces no impact whatsoever.

Putting  mod-squared terms   to
zero gives us the first-order Bogomolny equations, while the surface
term (the last one in Eq.~(\ref{bog})) gives the wall tension.
Assuming for definiteness that $\Delta m>0$ and choosing the upper sign
in (\ref{bog}) we get the BPS equations,
\begin{eqnarray}
\pz
\vp^A &=&
-\frac1{\sqrt{2}}\left(a_a\tau^a+a+
\sqrt{2}m_A\right)\vp^A,
\nonumber\\[3mm]
\label{wfoe}
\pz a^a &=&-
\frac{g_2^2}{2\sqrt{2}}\left(\bar{\varphi}_A\tau^a \vp^A
\right),
\nonumber\\[3mm]
\pz a &=&-
\frac{g_1^2}{2\sqrt{2}}\left(|\vp^A|^2
-2\xi
\right).
\end{eqnarray}

Tensions of the  walls satisfying the above equations are
given by the  surface term in Eq.~(\ref{bog}).\footnote{It is easy to check
that the very same result follows from the central charge (\ref{stenkaz}).}
Say, for the elementary walls $12\to 1B$ or
$12\to B2$, ($B=3,4$), this gives
\beq
T^{(12\to 1B)}_{\rm w}\,= T^{(12\to B2)}_{\rm w}\,=
\,   ( \Delta m )\,  \xi \,,
\label{ewten}
\eeq
where we use the fact that
$$
\Delta a \equiv (a)_{13} -(a)_{12} =\frac{m_1-m_3}{\sqrt 2}
=\frac{\Delta m}{\sqrt 2}\,.
$$
Remember, the elementary walls are those
 for which both vacua, initial and final,  have a common flavor.
 The wall $12\to 34$ can be considered as a
bound state of two elementary walls $12\to 1B$  and $1B\to CB$,
($C=3,4$, $B\neq C$). For the composite walls Eq.~(\ref{bog}) implies
\beq
T^{(12\to 34)}_{\rm w}\,=
 2 \, ( \Delta m )\,  \xi  \, ,
\label{bwten}
\eeq
since $\Delta a$ is twice larger.
We see that this wall has twice the tension of the
elementary walls. This means that the bound state is
marginally stable; the elementary BPS components
forming the composite wall do not interact. Equations (\ref{ewten})
and (\ref{bwten}) and the subsequent statement are valid
up to non-perturbative effects residing in the anomalous terms
in Eq.~(\ref{stenkaz}). For further discussion see Sect.~\ref{dotwvt}.

\subsection{Elementary domain walls}
\label{elwall}

It is time to explicitly  work out the solution to the first-order equations (\ref{wfoe})
for the domain wall interpolating between the vacua $(12)$ and $(1B)$ where
$B=3$ or 4. We assume that
\beq
m\,\gg\,\Delta m\,\gg\,\sqrt{\xi}=\sqrt{6\mu\,m}\, .
\label{xideltam}
\eeq
This condition allows us to find analytic domain  wall
solutions. In addition, it makes transparent the
physical reason for  the gauge field  localization   on domain
walls \cite{Shifman:2002jm}. Accepting (\ref{xideltam}) we guarantee,
as will be shown shortly, that the
quark fields (almost) vanish
inside the composite $(12\to 34)$-wall, to be treated in Sect.~\ref{1234}.
The only gauge symmetry
breaking surviving  inside this wall is   that induced by
the VEV of the SU(2) singlet adjoint field $a$.

Let us  choose the wall $12\to 14$ for definiteness.
The boundary conditions for the fields $a^3$ and $a$  are
obviously as follows (cf. Sects.~\ref{su2sym} and \ref{su2nsym}):
\beqn
a^3(-\infty) &=& 0, \qquad\qquad\,\,\,
a(-\infty) =-\sqrt{2}m_1 \,,\nonumber\\[4mm]
a^3(\infty) &=& -\frac1{\sqrt{2}}\Delta m, \qquad
a(\infty) =-\sqrt{2}m \,.
\label{121Bbca}
\eeqn

We see that the  range of variation of the fields
$a^3$ and $a$ inside the wall is of the order of $\Delta m$.
Minimization of  their kinetic energies
implies then that these fields are slowly varying.
Therefore, we may safely assume that the wall is thick on the scale of
$\xi^{-1/2}$;
the wall  size $R\gg 1/\sqrt{\xi}$.
This fact will be confirmed shortly, see also the previous
investigation \cite{Shifman:2002jm}.

\begin{figure}[h]
\epsfxsize=9cm
\centerline{\epsfbox{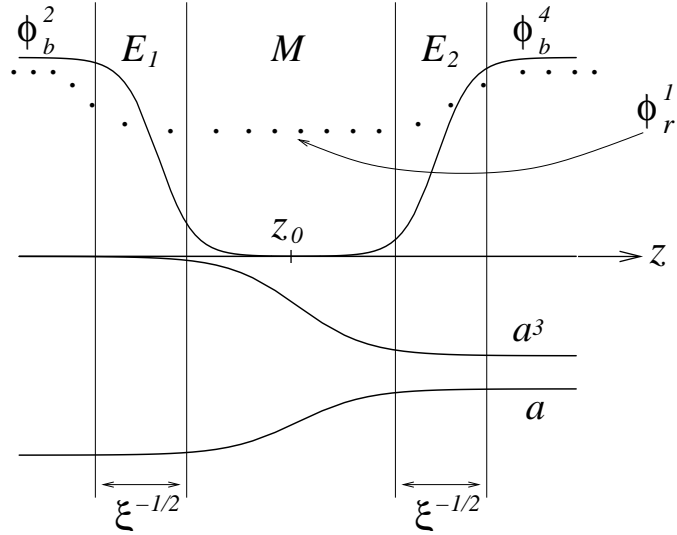}}
 \caption{Internal structure of the $12\to 14$ domain wall:
two edges (domains $E_{1,2}$) of the width $\sim \xi^{-1/2}$
are separated by a broad middle band (domain $M$) of the width $R$,
see Eq. (\ref{R}).}
\label{syfigthree}
\end{figure}

On the contrary, the quark fields vary inside small regions of the order of
$1/\sqrt{\xi}$ --- this scale  is determined by the masses of the light quarks
(\ref{r2mass}). In particular, $\vp_b^2$ varies from its VEV
in the 12-vacuum, (see Eq.~(\ref{qvev})),
$$
\langle \vp_b^2\rangle =\sqrt{\xi}\,,
$$
 at $z=-\infty$ to
zero near the the left edge of the wall (Fig.~\ref{syfigthree}), whereas
$\vp_b^4$ varies from zero to its VEV in the 14-vacuum,
$$
|\langle\vp_b^4\rangle|=\sqrt{\xi}\,,
$$
near the right edge of the wall. The $\vp_r^1$ quark field does not vanish
 inside the wall because it has a non-zero VEV
$$
\langle \vp_r^1\rangle=\sqrt{\xi}
$$
in both vacua, initial and final.
It acquires a constant value $\vp_{r0}^1$ inside the wall
which   will be determined shortly.

With these values of the quark fields inside the wall, the
 last  two equations in (\ref{wfoe}) tell us that the fields  $a^3$ and $a$ are
linear function of $z$ (cf. Ref.~\cite{Shifman:2002jm}). The solutions for
 $a^3$ and $a$ take the form
\beqn
a &=& -\sqrt{2}\left( m-\frac{\Delta m}{2}\,
\frac{z-z_0-R/2}{R}\right)\, ,
\nonumber\\[4mm]
a^3&=& -\frac{1}{\sqrt{2}}\, \Delta m\,
\frac{z-z_0+R/2}{R}\, ,
\label{a}
\eeqn
where the collective coordinate $z_0$ is
the position  of the wall center, while
$R$ is
the wall thickness  (Fig.~\ref{syfigthree}).
It is worth remembering that  $\Delta m$ is assumed positive.
The solution (\ref{a}) is valid in a wide domain of $z$,
\beq
\label{inside}
\left| z-z_0\right| < \frac{R}{2}\,,
\eeq
except narrow areas of size $\sim 1/\sqrt{\xi}$ near the edges of the wall
at $z-z_0=\pm R/2$.
Substituting the solution (\ref{a}) into the  last  two equations in
(\ref{wfoe}) we get
\beq
\label{R}
R=\frac{\Delta m}{ \xi}\left(\frac1{g^2_1}+\frac1{g^2_2}\right) .
\eeq
At the same time, the solution for $r$-quark inside the wall is
\beq
\vp_r^1=\vp_{r0}^1=\sqrt{\frac{2\xi}{g_2^2\left(\frac1{g^2_1}
+\frac1{g^2_2}
\right)}} \approx \sqrt{\frac{\xi}{2}}\, .
\label{r0}
\eeq
We see that  the $r$-quark field
inside the wall   differs from
 its  value in the bulk, generally speaking. Only if we take $g_1=g_2$ (which is {\em not} what comes out from the
SU(3)  ``protomodel," see Eq.~(\ref{normcond})) $\vp_{r0}^1$
 becomes equal to $\sqrt\xi$, its value in the bulk.
Since $\Delta m /\sqrt\xi \gg 1$, the result
(\ref{R}) shows that $R\gg1/\sqrt{\xi}$,  justifying our approximation.

As a test of the validity of  the solution above, let us verify that
 the solution $\vp^1_r$=const
 satisfies the first of   equations  (\ref{wfoe}) inside the wall.
Substituting solutions (\ref{a}) for the $a$ fields  in this equation
we get $\pz \vp_r=0$,  in full accord  with our solution (\ref{r0}).
Furthermore, we can now use the first relation in Eq.~(\ref{wfoe})
to determine the tails of the $b$-components of the 2,4-squark
fields inside the wall.

To this end, consider first the left edge (the domain $E_1$ in
 Fig.~\ref{syfigthree})
at $z-z_0=-R/2$. Substituting the above solution for $a$'s in the equation
 for $\vp_b^2$ we arrive at
\beq
\label{wq2}
\vp_b^2=\sqrt{\xi}\,
e^{-\frac{\Delta m}{2R}\left(z-z_0+\frac{R}{2}\right)^2}.
\eeq
This behavior is valid in the domain $M$, at  $(z-z_0+R/2)\gg1/\sqrt{\xi}$,
and shows that the field of the second quark flavor tends to zero
exponentially inside the wall, as was expected.

By the same token,  we can consider the behavior of
$b$-components of the fourth flavor squark field
near the right edge of the wall at  $z-z_0=R/2$. The
first of equation  in (\ref{wfoe})   for $A=4$
implies
\beq
\vp_b^4=e^{i\sigma} \,\sqrt{\xi}\,
e^{-\frac{\Delta m}{2R}\left(z-z_0-\frac{R}{2}\right)^2 }\,\,,
\label{wqB}
\eeq
which is valid in the
domain $M$ provided that
$$
(R/2-z+z_0)\gg1/\sqrt{\xi}\,.
$$
Here $\sigma$ is an Abelian wall modulus (of the phase type)
similar to that  discovered in our previous work \cite{Shifman:2002jm}
where the reader can find extensive explanations as to its origin.
Inside the wall
the fourth quark fields  tend to zero  exponentially too.

In the domains near the wall edges,
$$
z-z_0=\pm R/2\,,
$$
the fields $\vp_{r,b}^A$ as well as  $a^{3} $ and $a$
smoothly interpolate between their VEV's in the given vacua and the
inside-the-wall behavior  determined by Eqs.~(\ref{a}),
(\ref{wq2}), and (\ref{wqB}).
It is not difficult to check that these domains produce
contributions to the wall tension of the order of $\xi^{3/2}$, which makes
them negligible.

A comment is in order here regarding the collective coordinates
characterizing the elementary domain wall.
We have two collective coordinates in our wall solution:
the position of the center $z_0$ and the phase $\sigma$. In the effective
low-energy   theory on the wall world volume
they become (pseudo)scalar fields of the world volume (2+1)-dimensional theory,
$\zeta (t,x,y)$ and $\sigma (t,x,y)$, respectively.
 The target space of the second field is
$S_1$, as is obvious from Eqs. (\ref{wqB}).

In (2+1)-dimensional theory on the  wall the
compact (pseudo)scalar is equivalent to a U(1)  gauge field via the relation
\cite{Polyakov:1976fu}
\beq
F^{(2+1)}_{nm}={\rm const}\times
\ve_{nmk}
\, \pt^k\sigma\, ,
\label{F21}
\eeq
where $n,m,k=1,2,3$.

We see that our elementary domain wall localizes the  U(1)  gauge
field on its world volume, as was expected, and in full accord
with the string/D-brane notions.
 The physical reason for this localization was first
suggested in \cite{DS} and then elaborated in detail in \cite{Shifman:2002jm}
for the case of \ntwo QCD with the SU(2) gauge group (a model effectively reducible to SQED). In this particular aspect --- the gauge field localization
on the elementary wall ---
the present SU(2)$\times$U(1) model has  slight  distinctions
compared to that of Ref.~\cite{Shifman:2002jm} that are
worth mentioning.

\begin{figure}
\epsfxsize=6cm
\centerline{\epsfbox{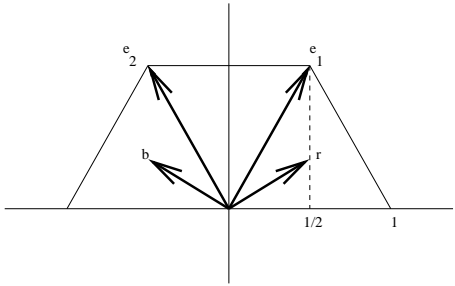}}
\caption{Root and weight vectors of the SU(3) algebra.}
\label{fi:su3}
\end{figure}

In the bulk the gauge symmetry is broken down to U(1)$^2$
by the VEV  of the $a^3$ adjoint field, and then, at a much lower scale,
it is completely broken  by the squark condensation. At the same time, inside the wall the only non-vanishing squark field is the
$r$-component of the first quark flavor.
Therefore, inside the wall the  U(1)  factor
orthogonal to the $r$-th weight vector of the  gauge group
SU(3) is restored. This  U(1)  factor is associated with $e_2$-root
of the gauge group, see Fig.~\ref{fi:su3}
where we imagine an  embedding of the SU(2)$\times$U(1)
gauge group in the SU(3) gauge group of our  underlying
``protomodel."  Thus we have a localization
of the $e_2$ gauge field on $12\to 14$ wall.

Note that this  field is dual to the one present in the bulk
\cite{DS,DV,Shifman:2002jm}. This means that if we put the $e_2$-monopole at
a certain point in the bulk, the $e_2$-string will be attached to this
monopole because monopoles are in the confining phase in the quark vacua
(see \cite{MY} and a brief review  in Sect.~\ref{naftin2} below on the flux tubes
in SU(2)$\times$U(1) \ntwo QCD).
As we will see in Sect.~\ref{naftin2}, it is  this string that will end on
our  elementary  walls, the string endpoint   playing the role of a (dual)
electric charge for the (2+1)-dimensional  U(1)  gauge field
(\ref{F21}) living
on the wall world volume.

In   conclusion of this  section it is worth noting that
the  scalar  $\zeta (x_n)$
and the gauge field $A_m(x_n)$ form the bosonic part of \ntwo vector
supermultiplet in 2+1 dimensions.

\subsection{Composite walls $(12\to 34)$ (bound states of
the type $(12\to 14) +(14\to 34)$}
\label{1234}

In this section we will consider the composite domain wall interpolating
between the vacua 12 and 34.

The boundary conditions
for all fields at  $z=-\infty$ are given by their VEV's in the 12-vacuum
\beqn
a^3(-\infty) &=& 0\,,
\nonumber\\[2mm]
 a(-\infty) &=& -\sqrt{2}\, m_1\,,
\nonumber\\[2mm]
\vp_r^1(-\infty) &=&
\vp_b^2(-\infty) =\sqrt{\xi}\, ,
\nonumber\\[2mm]
\vp_r^3(-\infty) &=& \vp_b^4(-\infty) =0\,,
\label{12bc}
\eeqn
while at $z=\infty$ they are given by VEV's in the 34-vacuum,
\beqn
a^3(\infty) &=& 0\, ,
\nonumber\\[2mm]
a(\infty) &=& -\sqrt{2}\, m_3\, ,
\nonumber\\[2mm]
\vp_r^1(\infty) &=& \vp_b^2(\infty) =0\, ,
\nonumber\\[2mm]
\vp_r^3(\infty) &=& \vp_b^4(\infty) =\sqrt{\xi}\, .
\label{12bcprime}
\eeqn
Now all quark fields (nearly) vanish inside the wall. The solution
for the $a$ fields in the middle domain M (Fig.~\ref{syfigthree})
is given by
\beqn
a
&=&
-\sqrt{2}\left( m_1-\Delta m\,
\frac{z-z_0+\tilde{R}/2}{\tilde{R}}\right),
\nonumber\\[3mm]
a^3 &=& 0\,,
\label{ac}
\eeqn
where we introduce the thickness $\tilde{R}$ of the composite wall,  to be
considered large, $\tilde{R}\gg 1/\sqrt{\xi}$, see below.
The equation for $a^3$ in (\ref{wfoe}) is trivially satisfied,
while the equation for $a$ yields
\beq
\tilde{R}=\frac{2\Delta m}{g_1^2\,  \xi}\,,
\label{tR}
\eeq
demonstrating that indeed  that $\tilde{R}\gg 1/\sqrt{\xi}$.
Note, that for a particular (unrealistic)
case $g_1=g_2$ (which we do {\em not} consider, since, according to Eq.~(\ref{normcond}),  $g_1\neq g_2$)
the size of the composite
wall is equal to that of the elementary ones, see (\ref{R}).

Substituting the above solutions in the first  two equations
in (\ref{wfoe}) we determine the fall-off of the quark
fields inside the wall. Namely, near the left edge
\beqn
\vp_r^1
&=&
\sqrt{\xi}\,
e^{-\frac{\Delta m}{2\tilde{R}}\left(z-z_0+\frac{\tilde{R}}
{2}\right)^2},
\nonumber\\[4mm]
\vp_b^2
&=&
\sqrt{\xi}\,
e^{-\frac{\Delta m}{2\tilde{R}}\left(z-z_0+\frac{\tilde{R}}
{2}\right)^2}\, ,
\label{wq12}
\eeqn
while near the right one
\beq
\vp^{kB}= \sqrt{\xi}\,(\tilde{U})^{kB}
e^{-\frac{\Delta m}{2\tilde{R}}\left(z-z_0+\frac{\tilde{R}}
{2}\right)^2}\,,\qquad B=3,4\, ,
\label{wq34}
\eeq
where the matrix $\tilde{U}$ is a matrix from the U(2) global flavor group,
which takes into account possible flavor rotations  inside the flavor pair
$B=3,4$. It can be represented as product of a U(1) a  phase factor and
a matrix  U  from  SU(2)
\beq
\label{flmat}
\tilde{U}=e^{i\sigma_0}\;U
\eeq
This matrix is parametrized by four phases, $\sigma_0$ plus
three phases residing in the matrix $U$.

The occurrence of these four wall moduli --- one related to U(1)
and three to SU(2) --- can be illustrated by the argument
which runs parallel to that outlined in Ref.~\cite{Shifman:2002jm}.
Indeed, in both vacua, 12 or 34, taken separately, one can always
use the symmetries of the theory to render the vacuum matrix
$\{\phi^{kA}\}$ diagonal,
\beq
\{\varphi^{kA}\}_{\rm vac} = \sqrt\xi \left(\begin{array}{cc}
1 & 0\\
0& 1
\end{array}
\right)\,,\qquad A = 1,2 \,\,\, \mbox{or} \,\,\,A = 3,4\,,
\eeq
with the real parameter $\sqrt\xi$ in front. When both 12-
and  34-vacua get involved
--- as is the case in the problem of the composite wall ---
a necessity arises of taking into account their relative alignment.
 The most concise way to see how these moduli emerge
 is through examination of a (non-local) {\em gauge-invariant} order parameter\,\footnote{The definition below is restricted to the {\em ansatz}
 (\ref{qqt}). In defining the non-local
 gauge invariant order parameter relevant to the domain walls
this is by no means  necessary. The general definition is
similar to that in Eq.~(\ref{nlgiop}) with the replacement
$\bar\varphi \to\tilde q$.
 }
 \beqn
{\cal O}_{A}^B(t,x,y)  &\equiv & \frac{1}{\xi}\left\langle
\bar\varphi_A (t,x,y \,|\,  z= - L) \, \exp \left\{i\,  \int_{-L}^L \, dz \left(
 \frac{\tau^a}{2} A_3^a (t,x,y \,|\,  z)\right.\right.\right.
 \nonumber\\[3mm]
&+&\left.\left.\left.
\frac{1}{2} A_3  (t,x,y \,|\,  z)\right) \right\}
 \varphi^B (t,x,y \,|\,  z=  L)\right\rangle\,,
 \label{nlgiop}
 \eeqn
where
$$A=1,2;\qquad B=3,4$$ and $L$ is a large parameter which we are
supposed to take to infinity at the very end (in practice, $L\gg\tilde R$.)

The order parameter ${\cal O}_{A}^B$ is non-singlet with respect to the
global U(2) inherent to our model upon its complete Higgsing.
In both vacua,   12  and  34, the order parameter ${\cal O}_{A}^B$ equals
to unit matrix ---
this is quite evident. However, is  {\em not} trivial on the
$12\to 34$ wall. On the wall ${\cal O}_{A}^B(t,x,y)$ reduces to a constant
U(2) matrix (independent of $t,x,y$) of the form
\beq
{\cal O}_{A}^B = \tilde U_A^B\,.
\label{ddmatrix}
\eeq
Applying all available symmetries of the model  at hand,
  the best we can do is to reduce
  the number of parameters residing in ${\cal O}_{A}^B$
  to four: one U(1) phase and three parameters of global SU(2).
There are no massless moduli in both vacua, initial
and final;  thus all of these four parameters  are collective
coordinates of the wall.

Below we will  identify these four moduli with (2+1)-dimensional gauge fields
living on the wall world volume   via duality relations
of the type presented in (\ref{F21}).

Thus, we get four gauge fields localized on the wall.
The physical interpretation of this result is as follows.   The quark fields are
condensed outside the $12  \to  34$ wall while inside they vanish.
This means that  dual gauge fields are severely confined
outside the wall while inside the confinement becomes inoperative.
This is precisely the mechanism of the gauge field
localization suggested in Ref.~\cite{DS}.

\subsection{Effective field theory on the wall}
\label{kinterms}

In this  section we work out the (2+1)-dimensional    low-energy
theory  of the moduli on the wall. First we will
discuss the elementary walls and then
 focus on the composite wall $12\to 34$.

\subsubsection{Elementary walls}
\label{elwo}

In this section we will deal with
 the elementary domain walls $1,2\to 1,B$ with $B=3$ or 4.
Our task is to work out the effective (2+1)-dimensional
theory for the wall collective coordinates (which become the world-volume
fields). For the elementary walls the overall situation is quite similar
to that discussed in our previous work \cite{Shifman:2002jm}.
Therefore, we will be rather fragmentary.

As was elucidated in Sect.~\ref{elwall}, the elementary wall has two
bosonic collective coordinates,
$z_0$ and  $\sigma$,  plus their fermionic counterparts
 $\eta^{\alpha f}$.
We make slowly varying fields dependent on
 $t,x,y \equiv x_n$  ($n=0,1,2$),
\beq
z_0 \to \zeta (x_n)\,,\quad \sigma \to \sigma  (x_n)\,,\quad
\eta^{\alpha f}\to \eta^{\alpha f}(x_n)\,.
\label{txyxn}
\eeq
We can limit ourselves to the
bosonic fields $\zeta (x_n)$ and  $\sigma  (x_n)$ ---
the  residual supersymmetry will allow us to
readily reconstruct the fermion part of the effective action.

The fields  $\zeta (x_n)$ and  $\sigma  (x_n)$ are in one-to-one
correspondence  with the  zero modes in the  wall background;
therefore,  they have no
potential terms in the world sheet theory, only kinetic. Our
immediate task  is to derive these
kinetic terms essentially repeating  the procedure of Ref. \cite{Shifman:2002jm}.
For $\zeta (x_n)$ this is trivial. Substituting
the wall solution (\ref{a}), (\ref{wq2}), and (\ref{wqB}) in  the action
(\ref{redqed}) and   accounting for the $x_n$ dependence of $\zeta (x_n)$,
with no further delay we arrive at
\beq
\label{kinz0}
\frac{T_{\rm w}}{2} \, \int d^3 x \; (\pt_n \zeta   )^2\, .
\eeq
This answer is quite general and would be valid for the translational modulus in
any model.

As far as the kinetic term for $\sigma  (x_n)$
is concerned an additional (albeit modest) effort is needed. We start from
Eq.  (\ref{wqB}) for the quark fields on the right edge of the wall
which depend on the phase $\sigma$
 parametrizing a relative  phase orientation of the fourth flavor
 with regards to the second one.
To calculate the corresponding  kinetic term
 we  have to modify our {\em ansatz}
for the gauge fields, namely,
\beqn
A^3_n
&=&
-\chi_3(z)\, \pt_{n}\sigma (x_n)\, ,
\nonumber\\[2mm]
A_n
&=&
\chi_0(z)\, \pt_{n}\sigma (x_n)\,.
\label{kingpot}
\eeqn
We    introduce  extra  profile functions  $\chi_0(z)$
and $\chi_3(z)$, much in the same way it was done in \cite{Shifman:2002jm}.
They have no role in    the static wall solution
{\em per se}. However,
in constructing the kinetic part of the world-volume theory
for the moduli fields their occurrence cannot be avoided.

These new profile functions
give rise to their own action, which must be subject to
minimization.  The gauge potentials (\ref{kingpot}) are  introduced
in order to cancel   the $x$ dependence of the quark fields
far from the wall (in the final quark vacuum at $z\to\infty$) emerging
through the  $x$ dependence of $\sigma (x_n)$, see  Eq. (\ref{wqB}).

Now let us turn to the kinetic terms in the (2+1)-dimensional
effective action coming from the quark kinetic terms in (\ref{redqed}).
For the first flavor we have
\beq
\label{dq1}
\nabla_n q^{r1}=-\frac{i}{2}(\pt_n\sigma )\,  (\chi_0 -\chi_3)\vp^{r1}.
\eeq
This expression is valid far away from the edges of the domain wall, that is to say,
in the middle domain $M$, where $q^{r1}$ is a non-vanishing constant (\ref{r0}),
and at $z\to \pm \infty$ where $q^{r1}$ tends to its vacuum expectation value $\sqrt{\xi}$.
To ensure the finiteness of the kinetic energy of the first quark flavor we
impose the following boundary conditions on the functions
$\chi_0$ and $\chi_3$:
\beq
\label{bc0-3}
\chi_0\to \chi_3, \; z\to\pm\infty.
\eeq
For the second flavor we have
\beq
\label{dq2}
\nabla_n q^{b2}=-\frac{i}{2} ( \pt_n\sigma ) (\chi_0 +\chi_3)\varphi(z_{-}),
\eeq
where we introduced the quark profile function
given by
\beq
\label{wallprofile}
\varphi (z)=
\left\{
\begin{array}{l}
\sqrt{\xi}\\[2mm]
\sqrt{\xi}\,
e^{-\frac{\Delta m}{2R}z^2},
\end{array}
\,\qquad
z\to\mp\infty \,,
\right.
\eeq
and the shorthand
\beq
\label{zpm}
z_{\pm}=z-z_0\mp\frac{R}{2}
\eeq
is implied. $z_{\pm}$
are the coordinates which vanish  at  the wall edges.

\vspace{1mm}

To make the kinetic energy of the second quark
flavor  finite   we impose the boundary conditions
\beq
\label{bc03-}
\chi_0\to \chi_3 \to 0, \qquad  z\to -\infty\,.
\eeq
A parallel  procedure for those quarks that have non-vanishing  VEV's in  the
final vacuum leads us to
\beq
\label{dqB}
\nabla_n q^{bB}=i ( \pt_n\sigma ) \left(1-\frac{\chi_0 +\chi_3}{2}\right)
\varphi(-z_{+}).
\eeq
This gives us the desired  boundary conditions for the functions $\chi$ at $z\to +\infty$,
\beq
\label{bc03+}
\chi_0\to \chi_3 \to 1, \qquad  z\to +\infty\, .
\eeq

Now we are ready to assemble all necessary elements.
Substituting (\ref{dq1}), (\ref{dq2}) and (\ref{dqB}) in  the action
and taking into account the kinetic term for the gauge fields
 we arrive at
\begin{eqnarray}
S_{2+1}^{\sigma}
&= & \int dz\,
\left\{\frac1{g_2^2}(\pz\chi_3)^2 + \frac1{g_1^2}(\pz\chi_0)^2
+\frac12(\chi_0-\chi_3)^2(\varphi^{r1} )^2
\right.
\nonumber \\[3mm]
 & + &
\left.
2\left(1-\frac{\chi_0+\chi_3}2\right)^2\varphi(-z_{+})^2 +
\frac12(\chi_0+\chi_3)^2\varphi(z_{-})^2
\right\}
\nonumber \\[3mm]
&\times& \int d^3 x \;\frac12(\pt_n \sigma )^2\,.
\label{kinsigint}
\end{eqnarray}
The expression in the integral over $z$ must be viewed
as an  action  for the $\chi$ profile  functions.
To get the classical solution for the BPS wall {\em and} the wall world-volume
 theory of the moduli  fields we  must  minimize this  action.
The minimization leads to two second-order equations for the functions
$\chi_0$ and $\chi_3$.
The solutions to these equations are linear in the middle domain $M$,
for both functions,
\beq
\chi_{0,3}= \frac{z-z_0+R/2}{R}\, .
\label{sfchiinm}
\eeq
Furthermore,  outside the domain wall the both functions exponentially approach
their boundary values (\ref{bc03+}), (\ref{bc03-}). This exponential approach is
controlled by the photon mass (\ref{mu1}) for the U(1) field and the W-boson mass
(\ref{m123}) for the  SU(2)  field (cf. \cite{Shifman:2002jm}).
Substituting the solution (\ref{sfchiinm})  in  the $\chi$
action  (\ref{kinsigint})  and
 taking into account (\ref{R}) we finally obtain
\beq
\label{kinsig}
S_{2+1}^{\sigma}=\frac{\xi}{\Delta m}\,
\int d^3 x \;\frac12 \, (\pt_n \sigma )^2\, .
\eeq

\vspace{2mm}

As has been already mentioned previously,  the compact scalar field
$\sigma (t,x,y)$
can be reinterpreted as a
 dual to the (2+1)-dimensional Abelian gauge field living on the wall,
see Eq.~(\ref{F21}). Note, that Eqs. (\ref{kingpot}) and
(\ref{sfchiinm}) demonstrate that the particular
combination of two  U(1)  gauge fields which is localized inside the wall is the combination with the following alignment:
$$
A^3_n=- A_n\,.
$$
This combination corresponds to the $e_2$-root of the SU(3) gauge
group of the ``prototheory," see Fig.~\ref{fi:su3}.
In particular, the $\vp^1$ quark which has a non-vanishing
$r$-component inside the wall is not charged under this combination.

\vspace{1mm}

The result presented in Eq.~(\ref{kinsig}) implies that the coupling constant of the
effective  U(1)  theory on the wall
\beq
\label{21coupling}
e^2_{2+1}=4\pi^2\,  \frac{\xi}{\Delta m}\, .
\eeq
This statement will help us make the definition
 of the (2+1)-dimensional gauge field outlined in Eq.~
(\ref{F21}) more precise,
\beq
\label{21gaugenorm}
F^{(2+1)}_{nm}=\frac{e^2_{2+1}}{2\pi} \,\varepsilon_{nmk}\,
\partial^k \sigma\, .
\eeq
As a result, the
effective low-energy theory of the moduli fields on the wall takes the form
\beq
\label{21theory}
S_{2+1}=\int d^3 x \, \left\{\frac{T_w}{2}\,\,  (\pt_n \zeta )^2+
\frac{1}{4\, e^2_{2+1}}\, [ F_{nm}^{(2+1)}]^2 +\mbox{fermion terms}
\right\}.
\eeq
The elementary wall at hand is  1/2 BPS-saturated --- it breaks four supercharges out of eight present in \ntwo theory. Thus we have
four fermion fields residing  on the wall,
$\eta^{\alpha f}$ ($\alpha, f = 1,2$). Because of the (2+1)-dimensional Lorentz  invariance of the on-the-wall theory
we are certain that these four fermion moduli fields form two
(two-component) Majorana spinors. Thus, the field content of the
world-volume theory we obtained is in full accord with
the representation of the (2+1)-dimensional extended
supersymmetry: one complex scalar field plus one Dirac
two-component fermion field.  Minimal  supersymmetry in
2+1 dimensions (with two supercharges) would require one
real scalar field and one Majorana two-component fermion field.
It is natural that we recover extended supersymmetry:
there are eight supercharges in our microscopic theory; the domain
wall at hand is 1/2 BPS; hence, we end up with four
supercharges in the world-volume theory.

\subsubsection{The composite wall}
\label{cowo}

Let us pass to the discussion of the effective world-volume theory
on the composite domain wall $12\to 34$. The emphasis will be put on novel elements appearing in  the theory of the
moduli fields on the composite wall which were absent in the case of
the elementary walls.

The first technical modification   compared to
Sect.~\ref{elwo} is that now we have
four independent compact moduli, rather than one ---
three residing  in the matrix $U$, Eq.~(\ref{flmat}) plus $\sigma_0$.
Therefore, in order to cancel the $x_n$-dependence of the
quark fields far away from the wall, in the final vacuum, we have to introduce in the {\em ansatz} four gauge fields,
\beqn
A_n
&=&
\chi_0(z)\, \pt_{n}\sigma^0(x_n)\, ,
\nonumber\\[3mm]
\frac{\tau^a}{2}\, A^a_n
&=&
-i\chi(z)\, \left[\pt_{n}U(x_n)\right]\, U^{-1}(x_n)\,.
\label{kingpotc}
\eeqn
Here  $\chi$ and $\chi_0$ are the  profile functions for  SU(2) and
U(1)  gauge fields, respectively. Calculations of the quark and
gauge kinetic terms run parallel to those in
Sect.~\ref{elwo} leading us to a key formula
\beqn
S_{2+1}^{\rm cm}
& =&
\int dz\,
\left[\frac1{g_1^2}(\pz\chi_0)^2
+(1-\chi_0)^2\varphi^{2}(-z_{+} )
+\chi_0^2\varphi^{2}(z_{-} ) \right]
\nonumber\\[3mm]
&\times &
\int d^3 x\;  \frac12(\pt_n \sigma_0 )^2
\nonumber\\[3mm]
& +&
\int dz\,
\left[\frac1{g_2^2}(\pz\chi)^2
+(1-\chi)^2\varphi^{2}(-z_{+} )
+\chi^2\varphi^{2}(z_{-} ) \right]
\nonumber\\[3mm]
&\times &
\int d^3 x\; {\rm Tr}\, \left[(U^{-1}\pt_n U)\;(U^{-1}\pt_n U)\right]\,,
\label{kinsigintc}
\eeqn
where the superscript cm stands for compact moduli.
The boundary conditions for the
functions $\chi$ and $\chi_0$ must be chosen
to ensure   finiteness  of energy in the domains
far away from the wall. This gives
\beqn
&& \chi_0\to \chi \to 0\,, \qquad  z\to -\infty\, ,
\nonumber\\[2mm]
&&\chi_0\to \chi \to 1\,, \qquad z\to +\infty\, .
\label{bcchic}
\eeqn
Equation (\ref{kinsigintc}) can be considered as an action functional for
 $\chi$ and $\chi_0$.

The functions $\chi$ and $\chi_0$ are determined by   minimization
of the above  action functional   which gives a second-order
equation for each function. We will not present them here,
since the reader can trivially get them himself/herself by
minimization. The solutions in the middle domain have
the already familiar linear form,
\beq
\chi_{0}=\chi= \frac{z-z_0+\tilde{R}/2}{\tilde{R}}\, ,
\label{sfchic}
\eeq
where the size of the composite wall $\tilde{R}$ is given in
Eq.~(\ref{tR}). Outside the wall the functions $\chi$ and $\chi_0$
exponentially approach   their boundary values (\ref{bcchic}).
The rate of approach is determined by the photon mass
(\ref{mu1}) for the function $\chi_0$,  while it is determined by the $W$-boson mass (\ref{m123}) for the function $\chi$.
Substituting the solution (\ref{sfchic}) back in  the action
(\ref{kinsigintc}) we obtain the following kinetic term:
\beq
S_{2+1}^{\rm cm} =\frac{\xi}{2\Delta m}\,
\int d^3 x\; \left\{\frac12(\pt_n \sigma_0 )^2
+\frac{g_1^2}{g_2^2}\,
{\rm Tr}
\left[(U^{-1}\pt_n U)\;
(U^{-1}\pt_n U)\right]
\right\}\,.
\label{kinsigc}
\eeq

\vspace{1mm}

Next, as in the elementary wall case,
we can try to dualize the moduli residing in $U$, as well as $\sigma_0$,
to convert them  in (2+1)-dimensional gauge fields
\beqn
F^{(2+1)}_{nm}
&=&
\frac{e^2_{2+1}}{2\pi} \,\varepsilon_{nmk}\,
\partial^k \sigma_0\, ,
\nonumber\\[3mm]
\frac{\tau^a}{2}\, F^{(2+1)a}_{nm}
&=&
-i\, \frac{g^2_{2+1}}{2\pi} \,\varepsilon_{nmk}\,
U^{-1}\pt_k U\, .
\label{21gaugefields}
\eeqn
Assembling all the above elements we obtain the action of
the world-volume effective theory,
\beqn
S_{2+1}
&=&
\int d^3 x \, \left\{\frac1{2e^2_{2+1}} (\pt_n a_{2+1} )^2+
\frac1{2g^2_{2+1}} (D_n a^a_{2+1} )^2 \right.
\nonumber\\[3mm]
&+&
\left. \frac{1}{4\, e^2_{2+1}}\, \left[ F_{nm}^{(2+1)}\right]^2
+\frac{1}{4\, g^2_{2+1}}\, \left[ F_{nm}^{(2+1)a}\right]^2
+\mbox{fermion terms} \right\},
\label{21theoryna}
\eeqn
of which a few comments are in order immediately.

The first comment refers to   four non-compact  $a,\, a^a$ moduli
which emerged in Eq.~(\ref{21theoryna}) seemingly out
of blue. We can use gauge transformation in the world volume theory
to put two of them to zero, say $a^{1,2}_{2+1}=0$. The other two
$a^3_{2+1}$ and $a_{2+1}$ should  be identified
with (linear combinations of) two centers of the elementary walls comprising
our composite wall.\footnote{Note that in Sect.~\ref{1234} we
worked out the solution for the composite wall as a bound state
of two elementary walls at zero separation. However,
in fact, this bound state is marginally unstable and has a zero mode
associated with the possibility of arbitrary separation between components.} More exactly, as $a_{2+1}$ has no
interactions whatsoever it is to be be identified with the center of mass
of the composite wall,
\beq
\label{centerid}
a_{2+1}=\sqrt{\xi\Delta m}\;e_{2+1}\;\frac1{\sqrt{2}}
\, (\zeta_1+\zeta_2)=
\pi\xi\;(\zeta_1+\zeta_2),
\eeq
while $a_{2+1}^3$ can be identified with the relative separation between
the elementary walls,
\beq
\label{separationid}
a_{2+1}^3=\sqrt{\xi\Delta m}\;g_{2+1}\;\frac1{\sqrt{2}}(\zeta_1-\zeta_2)
=\pi\xi\;\frac{g_1}{g_2}\;(\zeta_1-\zeta_2),
\eeq
where we use the fact that the  tension  of elementary walls is
 $T_w=\xi\Delta m$.

The second comment is devoted to a technical (but very important!)
element of the derivation of Eq.~(\ref{21theoryna}). In fact, this
world-volume action was obtained,
through the calculational procedure described above, only
at the quadratic level (i.e. omitting non-Abelian non-linearities).
This is rather obvious as our derivation, and in particular
the identification (\ref{21gaugefields}) and the effective action
(\ref{kinsigc}),
limits itself  to the quadratic   approximation.
Higher-than-quadratic terms in the (3+1)-dimensional action, such as the
commutator
term in the gauge field strength tensor, would produce four-derivative terms
in the (2+1)-dimensional theory (\ref{kinsigc}). Such
terms were explicitly omitted  in the  derivation above.
To  recover non-quadratic (truly non-Abelian) terms in Eq.~(\ref{21theoryna})
we use gauge invariance on the world volume.

The final remark is about the values of the
coupling constants in the  (2+1)-dimensional (``macroscopic")
theory in relation to the (3+1)-dimensional (``microscopic")
parameters. The  U(1) and SU(2) gauge coupling constants in (\ref{21theoryna})  are given by
\beqn
e^2_{2+1}
&=&
2\pi^2\,  \frac{\xi}{\Delta m}\, ,
\nonumber\\[3mm]
g^2_{2+1}
&=&
2\pi^2\,\,  \frac{g_1^2}{g_2^2}\,\frac{\xi}{\Delta m}\, .
\label{21couplings}
\eeqn

Our domain wall is a 1/2-BPS object so it preserves four supercharges
on its world volume. Thus, we must have the extended \ntwo supersymmetry,
with four supercharges, in the (2+1)-dimensional world-volume theory.
This is in accord with Eq.~(\ref{21theoryna}) in which
the U(1)  and  SU(2)  gauge fields are combined with the scalars
$a_{2+1}$ and $a^a_{2+1}$ to form the bosonic parts of \ntwo vector
multiplets.

Now let us discuss the possibility of spontaneous gauge symmetry
breaking in the world volume theory. Clearly if the adjoint
scalar $a^a_{2+1}$ develops a VEV, the SU(2) gauge symmetry is spontaneously broken in our world volume theory
(\ref{21theoryna}) on the composite wall.
We can always use gauge rotations to direct $a^a_{2+1}$ VEV along
third axis
in the color space, $\langle a^3_{2+1}\rangle \ne 0$. Then identification
(\ref{separationid}) shows that the separation $l=z_1-z_2$ between
two elementary walls which form our composite wall is nonvanishing.
In particularly the mass of the (2+1)-dimensional W-boson  is
given by the separation between elementary walls,
\beq
\label{21Wmass}
m_{W}^{2+1}=\pi\xi\;\frac{g_1}{g_2}\; l.
\eeq
The mass grows linearly with $l$. This is completely consistent with similar result for D-branes obtained in string theory.

Now let us discuss how  can one see this gauge symmetry breaking in
the (3+1)-dimensional bulk theory. Let us split our composite wall in
two elementary ones, say, 12 $\rightarrow$ 14 and 14 $\rightarrow$ 34.
Now we pull these two elementary walls apart making the
separation much larger than the wall thickness,
$l\gg R$. The separation $l$ must be much larger than  $R$ because the
scale $1/R$  plays a role of an ultraviolet cutoff in the
world-volume theory (\ref{21theoryna}).  Clearly two well separated elementary walls
have only two phase collective coordinates, one per each wall,
see Sect.~\ref{elwall}. If we dualize these phases we get two Abelian
gauge fields in the effective world volume theory. This corresponds to
breaking of SU(2)$\times$U(1) gauge symmetry down to
U(1)$\times$U(1) in  (\ref{21theoryna}). At separations
$l\gg R$ the masses of two (2+1)-dimensional W-bosons become
much larger than $\Delta m$
(see (\ref{21Wmass}) and they cannot be seen in the
effective low-energy world-volume theory.

One may ask where do  two extra ``non-Abelian"
phases of the composite wall disappear
at separations larger than the elementary wall width.
Of course, they do not disappear. They just pass into
Goldstone modes in the intermediate
14-vacuum.  Remember that the intermediate 14-vacuum has a Higgs
branch, see Sect.~\ref{su2nsym}. The two extra phases are now
associated with the massless moduli on this Higgs branch. At zero
separation these  phases are collective coordinates of the
composite wall. They belong to the (2+1)-dimensional world-volume
theory. At large separations they become bulk excitations
living in the intermediate vacuum. We will return to this issue
in Sect.~\ref{GSB} where we will show that only Abelian strings
can end on the composite wall when the separation between its
components gets larger than the thickness of the individual
components.

To conclude this section, we  reiterate that
the consideration presented above is 
certainly not  a ``rigorous derivation" of the non-Abelian
gauge invariance in the effective world-volume action 
(\ref{21theoryna}). Rather, it can be viewed as a motivated argument.
Our derivation   is carried out only at the 
quadratic level and does not take into account non-Abelian
nonlinearities. We identify four compact 
collective coordinates, to be dualized into four gauge 
fields living on the wall. We also calculate their kinetic terms 
which fix the values of the 3D gauge coupling constants.  
Direct calculation of cubic and quartic interaction terms, i.e. a {\em bona fide} complete derivation,
goes  beyond the scope of this  paper. This is a task  for   furture work.  

The gauge 
invariance in  (\ref{kinsigc}) is not apparent since (\ref{kinsigc})
is written in terms of {\em gauge invariant} phases. The gauge invariance 
of the world-volume theory appears 
only in Eq. (\ref{21theoryna}), after dualization.  
There are  quite compelling albeit indirect
arguments showing that our proposal (i.e. the SU(2)$\times$U(1) gauge 
theory (\ref{21theoryna})) is the correct generalization of 
(\ref{kinsigc}). First, 
the number of fields matches. We have four compact phases and two 
non-compact centers. Upon dualization, they fit into a vector 
multiplet of 3D ${\cal N} =2$ theory with the SU(2)$\times$U(1) gauge group. Say,
if the gauge group were U(1)$^4$, we would need four phases and
four non-compact coordinates, which we do not have.
Thus, the non-Abelian gauge symmetry of
the world-volume theory, in  effect,
is supported by supersymmetry.
Second, there are ony two distinct coupling constants in 
(\ref{kinsigc}), rather than four. This also indicates that three 
phases, upon dualization, should be unified in the SU(2) gauge 
theory.

\section{Non-Abelian flux tubes in \ntwo QCD}
\label{naftin2}
\setcounter{equation}{0}

In string theory gauge fields are localized on    D-branes because
fundamental open strings can end on   D branes
\cite{P}. In Ref. \cite{Shifman:2002jm}
we demonstrated that this picture is also valid in field theory, in the Abelian gauge field case. Namely, the Abelian flux tube was shown to end on the domain
wall. The reason for such behavior is easy to understand. In the Higgs
vacuum (in which electric charges condense), the  magnetic field is trapped
into flux tubes. However, inside the wall  quark fields (almost)
vanish. Therefore, the magnetic flux which is carried by the string
in the bulk can  spread over inside the wall.
The magnetic fields become electric upon dualization.
The string end-point
plays the role of the electric charge  for the gauge field localized on the wall
\cite{Shifman:2002jm}.

Our task   is to generalize this picture to cover the case of the
non-Abelian gauge fields. The main goal is finding a solution for a
1/4-BPS string-wall junction, in which a string carrying a non-Abelian
flux can end on the composite wall $ 12 \to 34$. We start implementation
of the string-wall junction program in earnest in Sects.~\ref{bsn2}
and \ref{swjunc}. Meanwhile, a brief introduction in
non-Abelian flux tubes will be in order, to acquaint the reader
with the necessary machinery. An advanced investigation
of non-Abelian flux tubes in various regimes will
be described elsewhere \cite{SYfut}.

Vortices in non-Abelian theories were studied in many papers
in recent years \cite{VS,HV,Su,SS,KB,KS,MY}. However,  in all
these examples of vortex solutions, the string
flux is always directed in the Cartan
subalgebra of the gauge group. This implies a (hidden) Abelian nature
of these strings. Clearly these strings cannot be used for our purposes
because their end-points on  domain walls cannot act as sources for
 non-Abelian fields.

Only recently  a special regime was found \cite{Auzzi:2003fs}
in which flux tubes acquire orientational zero modes which allow one
to freely
rotate the string flux inside a non-Abelian group.
This special regime is associated with
the presence of a certain combination of global gauge and flavor
symmetry   not broken by VEV's of scalar fields. Below we will show  that
precisely this regime is realized in 12- and 34-vacua of the theory under consideration.

The theory analyzed in Ref.~\cite{Auzzi:2003fs} is   \ntwo QCD
with the SU(3) gauge group  broken down to SU(2)$\times$ U(1), with four quark flavors, all  with the same mass. We review the string solution
found in this paper and adapt the analysis to our SU(2)$\times$U(1)
model. To begin with, however, we present some general arguments in a simple
toy model.

\subsection{How ``non-Abelian" are non-Abelian strings we deal with?}
\label{sprmo}

In Ref.~\cite{GS} it was proven that the only BPS-saturated
strings at weak coupling in ${\cal N}=1$ theories
are those of the ANO type, occurring in U(1) theories.
Here we speak of ``non-Abelian" BPS strings
in ${\cal N}=2$. A natural question to ask is
in which sense the BPS flux tubes under consideration are
Non-Abelian strings. A conceptual answer can be given in a
simplified model which does not even need to be supersymmetric.

Indeed, let us consider a (non-supersymmetric)
model which generalizes that of Abrikosov-Nielsen-Olesen,
and has two gauge groups, SU(2) and U(1),  and scalar fields of two flavors.\footnote{This model is also a version of the Higgs sector of the Standard model.} Denote two SU(2) doublet fields by
$\varphi^{(1)}_i$ and $\phi^{(2)}_j$, $\,\, i,j=1,2$. Then, introduce a
2$\times$2 matrix field
\beq
\Phi =\left(
\begin{array}{cc}
\phi^{(1)}_1 & \phi^{(2)}_1\\[2mm]
\phi^{(1)}_2 & \phi^{(2)}_2
\end{array}
\right)
\label{matrixphi}
\eeq
The covariant derivatives are defined in such a way that they act from
the {\em left},
\beq
\nabla_\mu \, \Phi \equiv  \left( \partial_\mu -\frac{i}{2}\; A_{\mu}
-i A^{a}_{\mu}\, \frac{\tau^a}{2}\right)\Phi\, ,
\label{dcde}
\eeq
We assume the action to have the form
\beqn
S&=&\int d^4x \left[\frac1{4g^2_2} \left(F^{a}_{\mu\nu}\right)^2 +
\frac1{4g^2_1}\left(F_{\mu\nu}\right)^2
\right.\nonumber\\[3mm]
&+& \left. {\rm Tr} (\nabla_\mu \Phi )^\dagger (\nabla_\mu \Phi ) +V(\Phi)
\right]\,,
\label{atm}
\eeqn
where, for the time being, the potential function $V$ is assumed to be gauge
invariant as well as invariant under the {\em global} U(2)
\beq
\Phi \to \Phi U_R\,.
\label{glo}
\eeq
Here $U_R$ is a constant matrix from U(2), and the multiplication is performed
from the {\em right}. The action (\ref{atm}) is invariant under the local U(2),
\beq
\Phi \to U_L (x) \Phi \,,
\label{lo}
\eeq
with   $A_\mu$ and $A_\mu^a$ transformed appropriately,
{\em and} under the global
U(2), Eq.~(\ref{glo}).

Models of the type (\ref{atm}) were engineered long ago  \cite{BarH}
with the purpose of  providing a set up for the spontaneous
breaking of the local (gauge) group $G$  down to a diagonal global $G$.
Indeed, with an appropriate choice of the
potential function $V$, one can ensure the vacuum expectation value
of  $\Phi$ to be diagonal,
\beq
\Phi_{\rm vac} = v\, \left( \begin{array}{cc}
 1 & 0\\[2mm]
0& 1
\end{array}
\right)\,, \qquad v\neq 0\,.
\label{vacphi}
\eeq
Then, since this vacuum is obviously invariant under the combined
multiplication
\beq
\Phi \to U_L \Phi U_R\,,
\label{combo}
\eeq
with $U_R= U_L^\dagger$,
the diagonal global U(2) will be preserved.

Now we can discuss topological defects of the string type.
Defects of the ANO type are always possible.
Indeed, put the SU(2) gauge field to zero (and temporarily forget
about it whatsoever). A non-trivial topology will be realized through
the U(1) winding of $\Phi$,
\beq
\Phi (x) = e^{i\alpha (x)}\,  v\,,\qquad |x|\to \infty\,,
\eeq
and
\beq
A_\ell = -2\, \varepsilon_{\ell k}\,  \frac{x_k}{r}\,,
\eeq
where $\alpha$ is the phase in the perpendicular
plane, and $r$ is the distance from the string axis in
the perpendicular
plane. Since $\pi_1 ({\rm U(1)})=Z$,
we will get in this way
 a   set of the ANO flux tubes with the arbitrary windings.

These are not the strings we are after, however. At first sight, the presence
of the SU(2) gauge symmetry, in addition to U(1),
does not create any new possibilities.
Indeed,  $\pi_1 ({\rm SU(2)})$ is trivial; one can readily unwind windings in SU(2) relevant to strings.

\begin{figure}
\epsfxsize=5cm
\centerline{\epsfbox{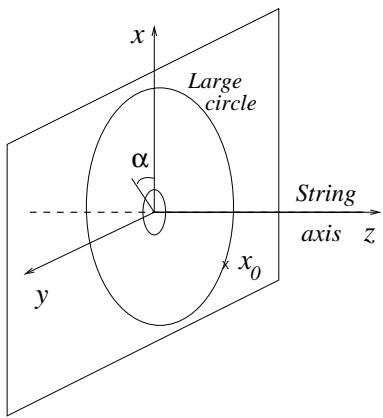}}
\caption{
Geometry of the string.}
\label{mmmon}
\end{figure}

\begin{figure}
\epsfxsize=9cm
\centerline{\epsfbox{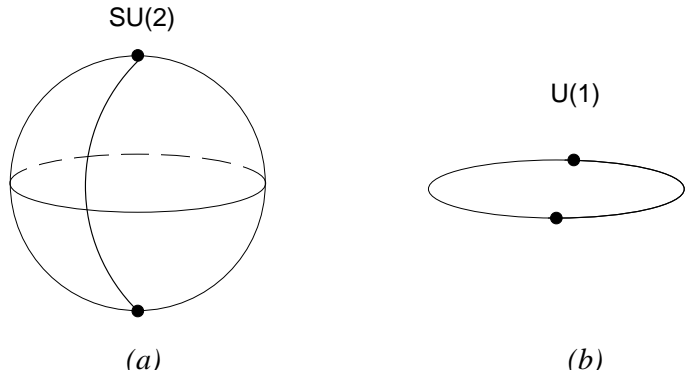}}
\caption{
Topology of the (1,0) and (0,1) strings.
The trajectories in the group spaces corresponding to
circumnavigating along the large circle in Fig.~\ref{mmmon}
are denoted by bold lines.}
\label{mmtue}
\end{figure}

Nevertheless, the fact that SU(2) has a center, $Z_2$, does create a new
possibility.\footnote{Of course, any element of U(1) can be considered as a center,
since this group is Abelian.} To see that this is the case, let us examine the following topology. A large circle in the plane perpendicular to the
string axis is depicted in Fig.~\ref{mmmon}
Assume that one starts from a certain point on this circle and makes a full
rotation around the string. Introduce the winding in SU(2), and assume
the full rotation above to bring us to the same element up to the center,
namely
\beq
\Phi (x) = e^{i\vec\omega (x)\vec\tau /2}\, \Phi (x_0)\to
 - \Phi (x_0)\quad{\rm at} \quad x\to x_0\quad{\rm after\,\,\, rot.},
\label{badc}
\eeq
see Fig.~\ref{mmtue}.
The condition (\ref{badc}) {\em per se} is
forbidden, since it results in a discontinuity of the
$\Phi$ field. One can eliminate this discontinuity
by supplementing the SU(2) winding above
by a U(1) winding  with the condition
\beq
\Phi (x) = e^{i \alpha (x) /2}\, \Phi (x_0)\to
 - \Phi (x_0)\quad{\rm at} \quad x\to x_0\quad{\rm after\,\,\, rot.},
\label{badcp}
\eeq
The formula
 \beq
 \Phi (x) = \exp\left( i \alpha (x) \frac{1\pm\tau^3}{2}\right)\, \Phi (x_0)\,,\quad
 \alpha (x) \to 2\pi
\quad{\rm after\,\,\, rot.},
\label{badcpp}
 \eeq
summarizes this pattern. Depending on the choice of the sign in
the exponent, only the $r$ or only the $b$ components of the
fields $\phi^{(1,2)}$ have a non-trivial winding.
It is clear from Fig.~\ref{mmtue} that one cannot unwind it.
It is also clear that the fluxes corresponding to the fields $A_\mu$ and $A_\mu^3$ are half the flux of the U(1) field of the ANO string.
We will refer to such strings as $(1,0)$ and $(0,1)$ --- the first
winding number corresponding
to the index $r$, the second to $b$. This notation
seems rather awkward given the way we introduced
the set up. It emerges naturally within the historical line of development, however;
see below. The standard ANO string emerges as a sum of the
$(1,0)$ and $(0,1)$ strings. The question as to which strings
are more favorable energetically depends on dynamical details.
We will return to it later.

A remarkable feature of the $(1,0)$ and $(0,1)$ strings is the appearance
of non-Abelian moduli which are absent in the ANO strings.
Indeed, while the vacuum field (\ref{vacphi}) is invariant
under the global SU(2) by virtue of Eq.~(\ref{combo}),
the string configuration (\ref{badcpp}) is not.
Therefore, if there is a solution of the form
(\ref{badcpp}) there is a family of solutions obtained from (\ref{badcpp}) by
the replacement
\beq
\Phi (x) \to \Omega \phi (x) \Omega^\dagger
\eeq
where $\Omega$ is an $x$-independent matrix
from U(2).  Generally speaking, it is parametrized
by four parameters. The U(1) factor is nothing but a
shift of the origin of the angle $\alpha$, however;
one should not count it. Thus, what remains is SU(2).
Moreover, in fact, it is SU(2)/U(1), as is clearly seen from
Eq.~(\ref{badcpp}). (Rotations around the third axis in the SU(2) space
leave the solution intact.) SU(2)/U(1) is the target
space of the $CP^1$ (or $O(3)$) sigma model which, thus,
provides the adequate description of the moduli
dynamics \cite{Hanany:2003hp,Auzzi:2003fs}.

It is just this aspect that allows us to refer to the strings above
as ``non-Abelian." They are as non-Abelian as it gets at weak coupling.

Note that the stability of the
$(1,0)$ and $(0,1)$ strings under consideration would be impossible
without the presence of the U(1) factor.

In conclusion, it is instructive to ask what happens if we explicitly
break the SU(2) flavor symmetry of the model
(\ref{atm}) by introducing unequal masses to the fields
$\phi^{(1)}$ and $\phi^{(2)}$, namely,
\beq
S_{\cal M} = \int d^4x\,{\rm Tr} (\Phi {\cal M}^2 \Phi^\dagger )\,,
\label{umte}
\eeq
where the mass matrix has the form
\beq
{\cal M}^2 = \left( \begin{array}{cc}
m_1^2 & 0\\[2mm]
0& m_2^2
\end{array}
\right)\,, \qquad m_1^2\neq m_2^2\,.
\label{vacphim}
\eeq
Intuitively it is clear that if $\Delta m^2$ is small,
the only change is as follows: the SU(2) symmetry of the
 $CP^1$ model must be broken, producing quasi-moduli from
the would-be moduli of    $CP^1$.  What survives is U(1), rather than SU(2).
Later, after re-introduction of supersymmetry,
we will see that this process is nothing but the transition from
the ${\cal N}=2$ $CP^1$ model to the one with the twisted
mass.\footnote{${\cal N}=2$ sigma models with twisted mass
were first constructed in Ref.~\cite{Alvarez}. The superspace/superfield
description was developed in Refs.~\cite{Gates:1983py,Gates:nk}.
In particular, the notion of a twisted  chiral superfield
was introduced in the second of these works. The word ``twisted''
appears for the first time in the given context in Ref.~\cite{Gates:nk}.}
Anticipating our further needs we present here
the bosonic part of the $CP^1$ model with the twisted mass,
\beq
{\cal L}_{\rm CP(1), t.m.}= G\, \left\{
\partial_\mu\bar\phi\partial^\mu\phi
-|\tilde{m} |^2 \bar\phi \phi
\right\} \,,
\label{bospatm}
\eeq
where  $G$ is the metric on the target space,
\beq
G  \equiv \frac{2}{g^2}\,\frac{1}{\left( 1+\phi\bar\phi\right)^2}\, ,
\label{13two}
\eeq
and $$\chi \equiv 1+\phi\bar\phi \,.$$ (It is useful to note that
the Ricci tensor $R = 2\, \chi^{-2}$.) The twisted mass
parameter $\tilde{m}$ introduced
in Eq.~(\ref{bospatm}) is related to the mass splitting
$\Delta m$ of the microscopic theory, $\tilde m = \Delta m$.

Anticipating further applications, we hasten to add that
the \ntwo superalgebra of the $CP^1$ model (which is our
macroscopic theory) is centrally extended,
namely,
\beq
\{Q_L \bar  Q_R\} = -i \, \tilde m q_{\rm U(1)}-  \tilde m \, \int
dz\, \partial_z\, h\, + \frac{ 1}{ \pi}
\int dz\, \partial_z \left(\chi^{-2}\,
\bar\Psi_R  \Psi_L \right) \,,\label{seventeen}
\eeq
where the first term is proportional to the U(1) charge $q_{\rm U(1)}$,
while in the second term $h = -\frac{2}{g^2}\,\, \frac{1}{\chi }$
presents the main impact of the twisted mass.
The last term is the central charge anomaly established
in Ref.~\cite{Losev:2003gs}.
It  is proportional to the difference between
the bifermion condensates in the final and initial
vacua. The central charge anomaly becomes important in the limit
$\tilde{m}\to 0$ corresponding to $m_1\to m_2$ in the microscopic theory. Then the classical terms $\sim q_{\rm U(1)}$ and $\sim h$
vanish, and the central charge is entirely determined by  the anomaly.

The above central charge is in one-to-one correspondence
with the BPS kinks in the $CP^1$ model. Sure enough, it must (and does) have a counterpart in the microscopic theory, see Sect.~\ref{lsmcc}.
Projecting them onto one another allows one
to establish relations between the parameters
of the microscopic and macroscopic theories \cite{SYfut},
for instance,
\beq
\frac{1}{g^2_{CP^1}}=\frac{2\pi}{g^2_2}\,.
\eeq
Let us note that $g^2_2$ in   runs according to the
formula of asymptotic freedom down to $\xi$,
(at $\Delta m =0$), where it is frozen in the bulk.
The asymptotic freedom running of $g^2_2$ is taken over and matched by that
of $g^2_{CP^1}$ in the macroscopic (world-volume) theory.

\subsection{Back to strings in ${\cal N}=2$}
\label{bsn2}

For definiteness let us consider strings in the 12-vacuum. To find
the BPS string
 solutions we use the same {\it ansatz} as in Eq.~(\ref{qqt}) and also
put adjoint fields, which are irrelevant for the string solutions, equal
to their VEV's (\ref{avev}). With these simplifications our theory
(\ref{redqed}) becomes
\beqn
S&=&\int d^4x \left[\frac1{4g^2_2} \left(F^{a}_{\mu\nu}\right)^2 +
\frac1{4g^2_1}\left(F_{\mu\nu}\right)^2
+ \left|\nabla_{\mu}
\vp^{A}\right|^2\right.\nonumber\\[4mm]
&+& \left.\frac{g^2_2}{8}\left(\bar{\vp}_A\tau^a \vp^A\right)^2+
\frac{g^2_1}{8}\left(\bar{\vp}_A \vp^A - 2\xi
 \right)^2
\right],
\label{le}
\eeqn
Clearly,  only those two flavors $A=1,2$ which develop
VEV's in the 12-vacuum
will play a role in the classical vortex solution. Other flavors remain  vanishing
on the solution. Hence, we  consider
the quark fields $\vp^{kA}$ to be $2\times 2$ matrices
in this section. Note, however, that the
additional two flavors are crucial in   quantum theory,  in
  keeping  the SU(2) interactions weak.

The string tension can be  written \`a la
Bogomolny \cite{B} as follows:
\begin{eqnarray}
T &=&\int{d}^2 x   \left\{
\left[\frac1{\sqrt{2}g_2}F^{*a}_{3} \pm
     \frac{g_2}{2\sqrt{2}}
\Big(\bar{\vp}_A\tau^a \vp^A\right)
 \right]^2+
\left[\frac1{\sqrt{2}g_1}F^{*}_{3} \pm
     \frac{g_1 }{2\sqrt{2}}
\left(|\vp^A|^2-2\xi \right)
\right]^2
\nonumber\\[4mm]
&+&
\left. \left|\nabla_1 \,\vp^A \pm
i\nabla_2\, \vp^A\right|^2
\pm
\xi  F^{*}_3
\right\}\,,
\label{bogs}
\end{eqnarray}
 where
\beq
 F^{*}_3 \equiv \frac12  \epsilon_{ij}  F_{ij }\,,\qquad (i,j=1,2),
\label{defstar}
 \eeq
plus the same for $ F^{*\,a}_3$,
are the coordinates in the plane orthogonal to the string axis directed
along the third (i.e. $z$) axis.
The Bogomolny representation implies the first-order equations for
the BPS strings,
\begin{eqnarray}
&&  F^{*a}_{3}+
     \frac{g^2_2}{2}\varepsilon
\left(\bar{\vp}_A\tau^a \vp^A\right)=0   , \qquad  a=1,2,3;
\nonumber \\[3mm]
&&   F^{*}_{3}+
     \frac{g^2_1}{2}\varepsilon
\left(|\vp^A|^2-2\xi \right)=0;
\nonumber \\[4mm]
 &&   ( \nabla_1  +i \varepsilon
\nabla_2)\, \vp^A=0, \qquad
\label{F38}
\end{eqnarray}
where $$\varepsilon = \pm 1$$ is the sign of the total flux specified below.

We first review the
U(1)$\times$U(1)  string solutions  found \cite{MY}  in
the unequal quark mass case,  and then show that in the limit of equal
quark masses additional orientational zero modes arise making the string
non-Abelian \cite{Auzzi:2003fs}.
For unequal quark masses some of the orientational moduli become quasi-moduli,
corresponding to passing from the $CP^1$ sigma model with no twisted mass
to that with a twisted mass, see Sect.~\ref{sprmo}.

The  U(1)$\times$U(1)  strings   can be    recognized, with no effort,  as particular
solutions of  Eqs.~(\ref{F38}).  To construct them we further restrict  the
gauge fields $A_{\mu}^a$ to a single  (third) color component $A_{\mu}^3$
(by setting $A_{\mu}^1 = A_{\mu}^2=0 $), and consider only the quark
fields  of the   $2 \times 2$     color-flavor diagonal form,
\beq
\label{ansatz}
\vp^{kA}(x)   \ne 0,  \qquad  {\rm for}
\quad   k=A   =1,2,
\eeq
with vanishing  other components.
For the unequal masses the relevant topological classification is
\beq
\label{pi1uu}
\pi_1\left(  \frac{{\rm U(1)}\times {\rm U(1)}}{ Z_2} \right)
=  Z^{2}\,,
\eeq
and   the  allowed  strings form  a lattice labeled by two integer
winding numbers.
To be more specific, assume that the first flavor winds $n$ times while
the second flavor winds $k$ times.  The solutions of Eq.~(\ref{F38}) are sought for using a ``natural" {\em ansatz},
\beqn
\vp^{kA}(x)
&=&
\left(
\begin{array}{cc}
  e^{ i \, n\,\alpha  }\phi_1(r) & 0  \\[2mm]
  0 &  e^{i \, k \,  \alpha }\phi_2(r) \\
  \end{array}\right)\,,
\nonumber\\[3mm]
A^3_{i}(x)
&=&
 -\varepsilon\, \epsilon_{ij}\,\frac{x_j}{r^2}\
\left[ (n-k)-f_3(r) \right]\, ,
\nonumber\\[3mm]
A_{i}(x)
&=&
- \varepsilon\, \epsilon_{ij}\,\frac{x_j}{r^2}\
\left[ (n+k)-f(r)\right]\,,
\label{sol}
\eeqn
where $(r, \alpha) $ are the polar coordinates in the (12)-plane while
the profile functions $\phi_1$, $\phi_2$ for the scalar fields and
$f_3$, $f$ for the gauge fields depend only on $r$ ($i,j=1,2$).

With this {\em ansatz},
the first-order equations (\ref{F38}) take the form \cite{MY}
\beqn
&&
r\frac{ d}{{ d}r}\,\phi_1 (r)- \frac12\left(f(r)
+  f_3(r)\right)\phi_1 (r) = 0\, ,
\nonumber\\[3mm]
&&
r\frac{ d}{{ d}r}\,\phi_2 (r)- \frac12\left(f(r)
-  f_3(r)\right)\phi_2 (r) =  0\, ,
\nonumber\\[3mm]
&&
-\frac1r\,\frac{ d}{{ d}r} f(r)+\frac{g^2_1}{2}\,
\left(\phi_1(r)^2 +\phi_2(r)^2-2\xi\right)= 0\, ,
\nonumber\\[3mm]
&&
-\frac1r\,\frac{ d}{{ d}r} f_3(r)+\frac{g^2_2}{2}\,
\left(\phi_1(r)^2 -\phi_2(r)^2\right)= 0\, .
\label{foest}
\eeqn
The profile functions in these equations
are determined by  the following boundary conditions:
\beqn
f_3(0)
&=&
 \varepsilon_{n,k}\left( n-k\right)\, ,
\qquad
f(0)=\varepsilon_{n,k}\,\left( n+k\right)\, ,
\nonumber\\[3mm]
f_3(\infty)
&=&
0\,,
\qquad \qquad\qquad\,\,\,
  f(\infty)=0
\label{fbc}
\eeqn
for the gauge fields.
The boundary conditions for  the quark fields are
\beqn
\phi_1 (\infty)
&=&
\sqrt{\xi}\,,\qquad   \phi_2 (\infty)=\sqrt{\xi}\,,
\nonumber\\[3mm]
\phi_1 (0)
&=&
0\, ,\qquad  \quad  \phi_2 (0)=0
\label{phibc}
\eeqn
for both $n$ and $k$ non-vanishing,
 while, say, for $k=0$ the only condition
at   $r=0$ is $\phi_1 (0)=0$.
Here the sign of the string flux is defined as
\beq
\label{sign}
\varepsilon = \varepsilon_{n,k} = \frac{n+ k}{|n+k|} =
{\rm sign}(n+ k) = \pm 1.
\eeq

The tension of a $(n,k)$-string is determined  by the
flux of the  U(1)  gauge field
alone    and is given by the formula \cite{MY,Auzzi:2003fs}
\beq
\label{ten}
T_{n,k}=\,2\pi \, \xi \, |n+ k| .
\eeq

Note    that  $ F^{*3}_3$ does not
enter the central charge of the \ntwo algebra and,
therefore,  does not affect  the string tension.
The stability of the string in this case is due to the  U(1)
factor of the SU(2)$\times$U(1)   group only.
Note also that $(1,0)$  and $(0,1)$-strings are exactly degenerate.

For a generic $(n,k)$-string equations (\ref{foest}) do not
reduce to the standard Bogomolny
equations. For instance,   for  the  (1,1)-string these equations reduce to
two Bogomolny equations while for the  (1,0)  and  (0,1)  strings
they do not. Numerical solutions for the two
``elementary''  (1,0) and  (0,1)  strings were obtained in
Ref.~\cite{Auzzi:2003fs}.

The charges of $(n,k)$-strings can be plotted on the Cartan plane of
the  SU(3)  algebra of the ``prototheory."
 We shall   use the convention of labeling
the flux of a given  string by the magnetic charge of the monopole which
produces this flux and can be attached to its end.
 This is possible, since both, string fluxes and the monopole
charges, are elements of the group $\pi_1({\rm U(1)}^2) =
{Z}^{2}$. This convention is quite convenient because specifying
the flux of a given string automatically fixes the charge of
the monopole that it confines.

\begin{figure}
\epsfxsize=7cm
\centerline{\epsfbox{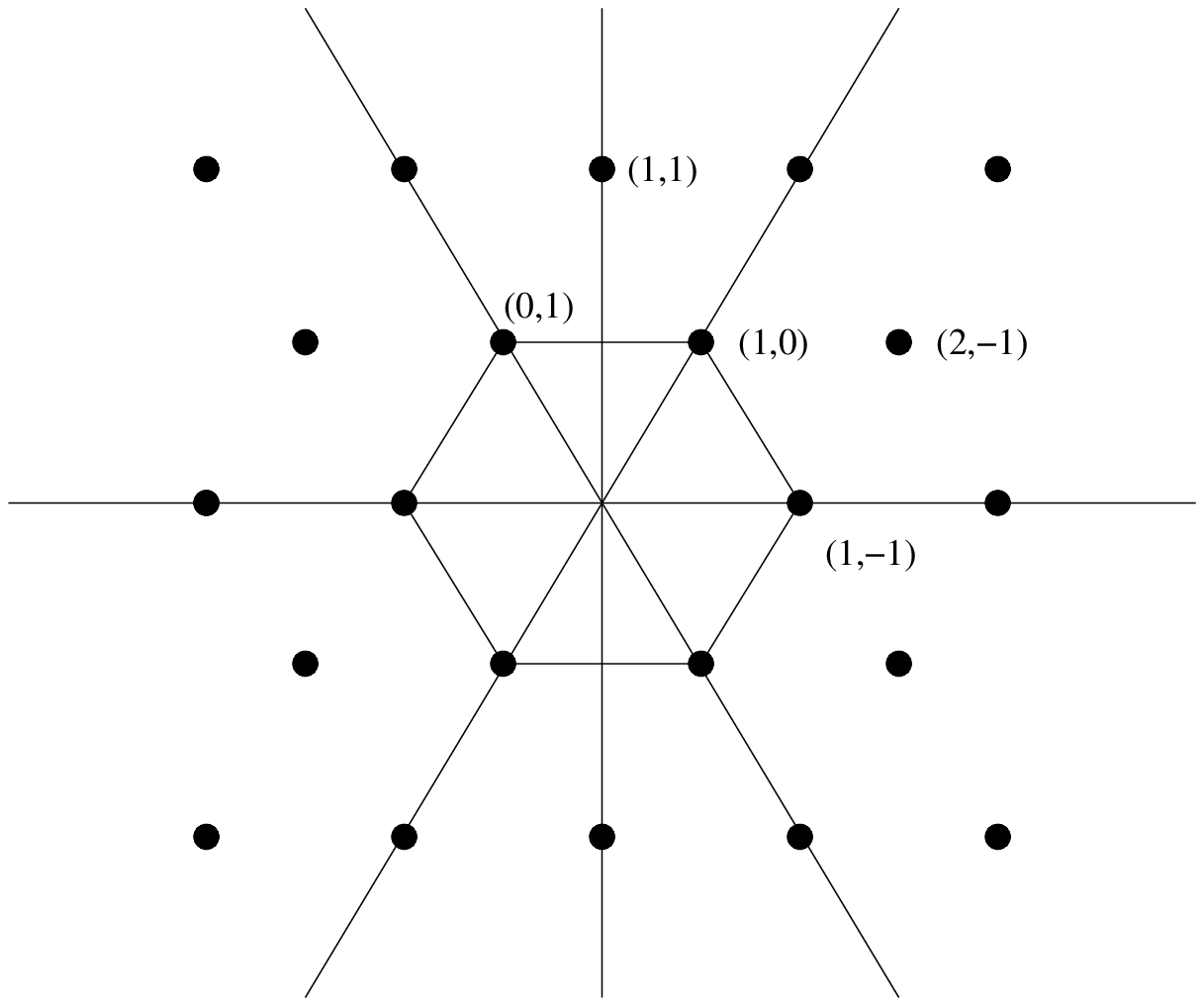}}
\caption{ Lattice of $(n,k)$   vortices.    }
\label{fi:lattice}
\end{figure}

Our strings are formed as a result of  the quark condensation; the quarks
have electric charges equal to the  weights of the SU(3)
 algebra.  The Dirac quantization condition tells us  \cite{MY} that
the lattice of  the $(n,k)$-strings is
 formed  by roots of the   SU(3)  algebra.
The lattice of  $(n,k)$-strings is shown in
Fig.~\ref{fi:lattice}. Two strings,  $(1,0)$ and $(0,1)$, are the
``elementary'' or ``minimal."   They are BPS-saturated. All other strings can be
 considered as   bound states of these ``elementary'' strings.
 If we plot two lines
along the charges of these
``elementary'' strings (Fig.~\ref{fi:lattice}) they divide the
lattice into four sectors. It turns out \cite{MY} that the strings
in the upper and lower sectors
are BPS but they are marginally
unstable. At the same time, the strings in
the right and left sectors
are (meta)stable bound states of the ``elementary''
ones but they are not BPS.

Now, let us generalize the string solutions (\ref{sol}) to the case
of the equal quark masses,  when the SU(2)$\times$U(1) gauge group is not broken
by the difference of the quark masses, as is the case in the 12-vacuum.
 The relevant  homotopy group in this case is the fundamental group
\beq
\label{pi1suu}
\pi_1\left( \frac{{\rm SU(2)}\times {\rm U(1)}}{ Z_2} \right)=  Z\,,
\eeq
replacing Eq.~(\ref{pi1uu}).
This means that the lattice of $(n,k)$-strings reduces to a tower
labeled by one integer $(n+k)$.
For instance, the $(1,-1)$-string becomes
completely unstable.  On  the restored  SU(2) -group manifold
it corresponds to a winding along the
meridian on the sphere $S_3$.
Clearly this winding can be shrunk to nothing
by contracting the loop towards the
north or south poles \cite{Shifman:2002yi}.

On the other hand, the
$(1,0)$ and $(0,1)$ strings cannot be shrunk because their winding is
  half a circle (Fig.~\ref{mmtue}). They have
the same tension
\beq
\label{miniten}
T_{1}=\,2\pi \xi
\eeq
 for equal quark masses and,
thus, apparently  belong to a doublet of   SU(2) .

Below we will show that there is
a continuous deformation of the $(1,0)$-string solution transforming
it into a $(0,1)$-string. This deformation leaves the string tension
unchanged and,  therefore,  corresponds to an orientational  zero mode
\cite{Auzzi:2003fs}.

First  let us fix the unitary  gauge (at least globally,
which is enough for our
purposes) by imposing the condition that the
squark VEV's are given precisely by (\ref{qvev}), and so all gauge
phases vanish.
Now transform  the $(1,0)$-string solution (\ref{sol}) into the
unitary gauge, which corresponds to the singular gauge, in which the
string flux comes from the singularity of the gauge potential at zero.
In this gauge the solution (\ref{sol}) for the $(1,0)$-string
takes the form
\beqn
\vp^{kA}
&=&
\left(
\begin{array}{cc}
  \phi_1(r) & 0  \\[2mm]
  0 &  \phi_2(r) \\
  \end{array}\right),
\nonumber\\[4mm]
A^3_{i}(x)
&= &
\epsilon_{ij}\,\frac{x_j}{r^2}\,
f_3(r),
\qquad
A_{i}(x) =  \epsilon_{ij}\,\frac{x_j}{r^2}\,
f(r)\, .
\label{s10}
\eeqn
Now, please,  observe   that  a global diagonal subgroup in the product
of gauge and flavor symmetries SU(2)$_{C}\times$SU(2)$_{F}$     is
not broken by the quark VEV's. Namely,
\beq
\label{cf}
{\rm U}_{C+F}\, \langle q\rangle\, {\rm U}^{-1}_{C+F}=\langle q\rangle\,,
\eeq
where U$_{C+F}$ is a global rotation in  SU(2)  while the quark VEV matrix
is given by Eq.~(\ref{qvev}). We refer to this unbroken group as  SU(2)$_{C+F}$.

Let us apply this global rotation to the $(1,0)$ string solution
(\ref{s10}). We find
\beqn
\vp^{kA}
&=&
{\rm U}_{C+F}\left(
\begin{array}{cc}
  \phi_1(r) & 0  \\[2mm]
  0 &  \phi_2(r) \\
  \end{array}\right){\rm U}^{-1}_{C+F},
\nonumber\\[3mm]
\frac{\tau^a}{2}\,
A^a_{i}(x)
&=&
\frac12\,n^a\tau^a\epsilon_{ij}\,\frac{x_j}{r^2}\,
f_3(r)\,,
\nonumber\\[4mm]
A_{i}(x)
&=&
 \epsilon_{ij}\,\frac{x_j}{r^2}\,
f(r)\, ,
\label{sna}
\eeqn
where we define
\beq
\label{na}
{\rm U}_{C+F}\tau^3\,  {\rm U}^{-1}_{C+F}=n^a\tau^a\, .
\eeq
Here  $n^a$ is a unit vector on $S_2\,$, $\vec n^2=1$.

It is easy to check that the rotated string (\ref{sna}) is a solution
to the non-Abelian first-order equations (\ref{F38}).
Clearly the   solution (\ref{sna}) interpolates between $(1,0)$ and $(0,1)$
strings. In particular it gives a $(1,0)$-string for $n=(0,0,1)$ and
a $(0,1)$-string for $n=(0,0,-1)$.

The vector $\vec n$ has a clear-cut physical meaning.
Its orientation is the orientation of the magnetic flux.
The construction above --- which was carried out in the singular gauge ---
shows that  the  SU(2)  flux of the string is directed along
the vector $n^a$.
This fact becomes even more transparent, if  we examine a gauge-invariant
definition of the magnetic flux of the non-Abelian string, which is
very instructive. This can be done as follows.
Define  ``non-Abelian" field strength (to be denoted by bold letters)
as follows:
\beq
{\bf{F}}_3^{*a} =\frac{1}{\xi}\,{\rm Tr}
\left(\Phi^\dagger F_3^{*b}\frac{\tau^b}{2}\Phi\, \tau^a
\right)
\eeq
From the very definition it is clear that this field
is {\em gauge invariant}. Moreover, it is clear
from (\ref{sna}) that
\beq
{\bf{F}}_3^{*a} =- n^a\frac{(\phi_1^2+\phi_2^2)}{2\xi}\frac1r\frac{df_3}{dr}\,.
\label{ginvF}
\eeq
Thus, the physical meaning of these moduli is as follows.
The flux of the color-magnetic field in the flux tube
 is directed along $\vec n$.
We see that  the SU(2)$_{C+F}$  symmetry is physical and does not
correspond to any of the gauge rotations which are ``eaten up"
by the Higgs mechanism. At the same time, a non-Abelian gauge
group --- a ``new color" --- is resurrected.
For strings in Eq.~(\ref{sol}) the ``new-color"-magnetic flux is
directed along the third axis in the O(3) group space, either
upward or downward.

The  SU(2)$_{C+F}$  symmetry is exact and the tension of the string
solution (\ref{sna}) is independent of   $n^a$,
see (\ref{miniten}).    However,  an explicit vortex solution breaks
the exact SU(2)$_{C+F}$ in the following manner:
\beq
{\rm SU}(2)_{C+F} \to {\rm U}(1)\,.
\eeq
Two angles associated with vector $n^a$ becomes two
orientational bosonic zero modes of the string. The vector
$n^a$  parametrize the
quotient space SU(2)/U(1)$ \sim  CP^1 \sim S^2$. This means that,
as we have already explained in Sect.~\ref{sprmo},  the
(1+1)-dimensional
low-energy effective theory for these orientational zero modes is
the O(3) sigma model (O(3) sigma model is the same as $CP^1$ sigma model;
if we started from SQCD with the gauge group
SU(N)$\times$U(1),
we would instead arrive  \cite{Hanany:2003hp,Auzzi:2003fs} at $ CP^{N-1}$).
Since the string is 1/2-BPS saturated we have four supercharges
in the effective world sheet theory. This corresponds to \ntwo
supersymmetry in (1+1)-dimensions.

Classically the O(3) sigma model is characterized by
a spontaneous breaking of the O(3) symmetry  leading to
two massless Goldstone bosons. This is to say  that in the
quasiclassical treatment the vector $n^a$   points in some
particular direction  for a given string.

However,   quantum physics of \ntwo sigma model is quite
different. The model is asymptotically free and runs into a
strong coupling regime at low energies.
This theory has a dynamically generated
mass gap
\beq
\Lambda_{CP^1}\sim \sqrt{\xi}\exp{(-2\pi /g_{CP^1}^2)} \sim \sqrt{\xi}\exp{(-4\pi^2/g_2^2)}\,.
\label{sigmascale}
\eeq
There is no spontaneous breaking of O(3),
and no Goldstone bosons are generated. In terms of strings in four dimensions
this means that the string orientation vector $n^a$
has no particular direction.
The O(3) sigma model has two vacua \cite{HoVa}.
In the microscopic  four-dimensional picture this means that we have two
``elementary" non-Abelian
strings which form a doublet with respect to SU(2)$_{C+F}$.

Note however, that  they are {\em not} the $(1,0)$
and $(0,1)$ strings of the quasiclassical
U(1)$\times$U(1) theory.
 In both strings the vector $n^a$ has no particular
direction. Still the number of ``elementary'' string states
remains the same --- two ---  in the limit of equal quark masses.

The O(3) sigma model has a kink interpolating between the two vacua.
In four dimensions this interpolation will be  interpreted as a monopole
which produces a junction of two ``elementary" non-Abelian
strings \cite{Auzzi:2003fs,Tong:2003pz}. This monopole lives on
the string world sheet  because monopoles are in the confining
phase in our theory, and do not exist as   free states.

The charge of this monopole lies entirely inside the SU(2) factor
of the gauge group. If $\Delta m\neq 0$,  its charge is $(1,-1)$.
Classically the mass of this monopole is  $\Delta m \, (4\pi/g^2_2)$
and tends to zero when the gauge symmetry is enhanced from U(1)$\times$U(1) to SU(2)$\times$U(1) at $\Delta m =0$.
Simultaneously its size becomes infinite (cf. \cite{We}).
However, in quantum theory
the story is different. This
monopole has a finite size  since there are no massless states in
the O(3) sigma model. It is massive  but extremely light
with a mass determined by the scale  of the sigma model
$\Lambda_{CP^1}$, see Eq.~(\ref{sigmascale}).
The mass of this monopole is lifted from zero
and is given by the anomalous term
 in the central charge (\ref{seventeen}) of the O(3) sigma model.
Its charge is no longer $(1,-1)$ because it
interpolates now between quantum vacua of the O(3) sigma model
for which the vector $n^a$ has no particular direction.
Further details are reported in \cite{SYfut}.

\section{String-wall junctions}
\label{swjunc}
\setcounter{equation}{0}

In this section we  derive  the BPS equations and find a
1/4-BPS solutions for  string-wall junctions.
First we work out the first-order equations for string-wall
junctions then find a solution of Abelian string ending on the
elementary wall and, finally, discuss a non-Abelian string ending on
the composite wall.

\subsection{First-order equations for junctions}
\label{stringending}

In Ref.~\cite{Shifman:2002jm} we found   string-wall junction  solution
picking up   two supercharges (1/4 BPS!)
 which  act trivially both
on  the string and wall solutions.
Here we take a slightly different route inspecting the  Bogomolny
representation  for the energy functional. We keep the quark, adjoint
and gauge fields in our action because all of these fields
play a role in the string-wall junction.

It is natural to assume that at large separations from the string
end-point  at $r=0$, $z=0$,
the wall is almost parallel to the $(x_1,x_2)$ plane while the string
is stretched along the $z$ axis at negative $z$.
 Since
both solutions, for the string and the wall,  were obtained using the
{\em ansatz} (\ref{qqt}) we restrict our search for  the wall-string junction
to the same {\em ansatz}. As usual, we look for a static solution
assuming that  all relevant fields can depend only on $x_n$, ($n=1,2,3$).

Then we can represent the energy functional of our theory (\ref{redqed})
as follows:
\beqn
E
&=&
\int{d}^3 x   \left\{
\left[\frac1{\sqrt{2}g_2}F^{*a}_{3} +
     \frac{g_2}{2\sqrt{2}}
\left(\bar{\vp}_A\tau^a \vp^A\right)
 + \frac1{g_2}D_3 a^a\right]^2
\right.
\nonumber\\[4mm]
&+&
\left[\frac1{\sqrt{2}g_1}F^{*}_{3} +
     \frac{g_1 }{2\sqrt{2}}
\left(|\vp^A|^2-2\xi \right)
+ \frac1{g_1}\pt_3 a\right]^2
\nonumber\\[4mm]
&+&
\frac1{g_2^2}\left[\frac1{\sqrt{2}}(F_1^{*a}+iF_2^{*a})+(D_1+iD_2)a^a
\right]^2
\nonumber\\[4mm]
&+&
\frac1{g_1^2}\left[\frac1{\sqrt{2}}(F_1^{*}+iF_2^{*})+(\pt_1+i\pt_2)a
\right]^2
\nonumber\\[4mm]
&+&
\left. \left|\nabla_1 \,\vp^A +
i\nabla_2\, \vp^A\right|^2+
\left|\nabla_3 \vp^{A}+\frac1{\sqrt{2}}\left(a^a\tau^a +a +\sqrt{2}m_{A}
\right)\vp^A\right|^2\right\}
\nonumber\\[4mm]
&+&
 \mbox{ surface terms},
\label{bogj}
\eeqn
where we assume that the quark mass terms and adjoint fields are real.
The surface terms are
\beq
E_{\rm surface}=
\xi  \int d^3 x F^{*}_3
+\sqrt{2}\xi
\left.
\int d^2 x \langle a\rangle\right|^{z=\infty}_{z=-\infty}
- \sqrt{2}\, \frac{\langle a^3\rangle}{g_2^2}\, \int d S_n
F^{*3}_{n}\, ,
\label{surface}
\eeq
where the integral in the
last term runs  over a large two-dimensional sphere at $x^2_n\to
\infty$, and \beq F^{*\,3}_n \equiv \frac12  \epsilon_{nij}
F_{ij}^3\,, \label{defstarp} \eeq cf. Eq.~(\ref{defstar}). This
is in full accord with the general discussion in
Sect.~\ref{ccn2}.

The Bogomolny representation (\ref{bogj})
 leads us to the following first-order equations:
\begin{eqnarray}
&& F^{*}_1+iF^{*}_2 + \sqrt{2}(\pt_1+i\pt_2)a=0\, ,
\nonumber\\[4mm]
&& F^{*a}_1+iF^{*a}_2 + \sqrt{2}(D_1+iD_2)a^a=0\, ,
\nonumber\\[3mm]
&& F^{*}_{3}+\frac{g_1^2}{2} \left(\left|
\varphi^{A}\right|^2-2\xi\right) +\sqrt{2}\, \pt_3 a =0\, ,
\nonumber\\[3mm]
&& F^{*a}_{3}+\frac{g_2^2}{2} \left(\bar{\vp}_{A}\tau^a \varphi^{A}\right)
+\sqrt{2}\, D_3 a^a =0\, ,
\nonumber\\[3mm]
&& \nabla_3 \vp^A =-\frac1{\sqrt{2}}\left(a^a\tau^a+a+\sqrt{2}m_A\right)
\vp^A\, ,
\nonumber\\[4mm]
&& (\nabla_1+i\nabla_2)\varphi^A=0\, .
\label{foej}
\end{eqnarray}
These are our {\em master equations}.

Once these equations are satisfied the energy of the BPS object is given by
Eq.~(\ref{surface}). Please, observe that Eq.~(\ref{surface}) has three terms
corresponding to central charges of the string, domain wall and monopoles
of the SU(2) subgroup,
respectively. Say, for the string the three-dimensional integral in
the first term in Eq.~(\ref{surface})
gives the string length   times its flux. For the wall the two-dimensional
integral in the second term in (\ref{surface}) gives the area of a wall
times the tension.
For the monopole the integral in the last term in (\ref{surface})
 gives the monopole flux.
  This means that our master equations (\ref{foej}) can be used to study
the BPS strings, domain walls,  monopoles, and all their possible junctions.

It is instructive to  check that  the wall,   string and   monopole
solutions, separately,
satisfy these equations. Say, we start from the wall solution.
In this case the gauge fields are put to zero,
and all fields depend only on $z$. Thus, the first two
and the last two equations in (\ref{foej}) are trivially satisfied.
The components of  the gauge fields $F^{*}_3$ and  $F^{*a}_3$
 vanish in the third and fourth  equations; hence
 these equations reduce to the last  two equations in Eq.~(\ref{wfoe}). The
 fifth equation in (\ref{foej}) coincides  with  the first one
in (\ref{wfoe}).

For the string which lies, say, in the  12-vacuum, all quark
fields vanish except
 $q^A$, $A=1,2$ while $a$ and $a^a$ are  given by their VEV's.
The gauge flux is directed along the $z$ axis, so
that $ F^{*}_1=F^{*}_2= F^{*a}_1=F^{*a}_2=0$.
All fields depend only on the  coordinates $x_1$ and $x_2$. Then the first
two equations and
the fifth  one  in (\ref{foej}) are trivially satisfied. The third and
the fourth  equation
reduce to the first two ones in Eq.~(\ref{F38}). The last equation
in (\ref{foej})
for $A=1,2$ reduces to the last equation in (\ref{F38}), while for $B=3,4$
these equations are trivially satisfied.

Equations for the monopole arise from the ones in (\ref{foej})
in the limit $\xi=0$. Then all quark fields vanish, and
Eq.~(\ref{foej}) reduces to the
standard first-order equation for the monopole in the BPS limit,
\beq
\label{monopole}
F^{*a}_n + \sqrt{2}D_n a^a=0,
\eeq
while $a$ is given by its VEV and the U(1) gauge field vanishes.

In particular, Eq. (\ref{surface}) shows that the central charge
of the SU(2) monopole is
determined by $\langle a^3\rangle$ which is proportional
to the quark mass difference
in the given vacuum. Say, for the monopole in  the 12-vacuum  it gives  zero.  However, as was mentioned
at the end of Sect.~\ref{bsn2},  the mass of this  monopole
is lifted from zero at   $\xi\neq 0$. In this case this monopole becomes
a junction of two ``elementary'' strings of
the SU(2)$\times$U(1) theory and acquires
a nonvanishing mass due to non-perturbative effects in the O(3) sigma
model on the string world sheet.

Let us note that the Abelian version of the first-order master equations
(\ref{foej}) was first derived in Ref. \cite{Shifman:2002jm} and
used to find a 1/4-BPS solution for the string-wall junction. Quite
recently a non-Abelian version for SU(2)$\times$ U(1) theory was
used \cite{Tong:2003pz} to study the junction of two
``elementary'' strings via a small-size monopole  at $\Delta m \neq 0$
and large.

\subsection{The Abelian  string
ending on the elementary wall}
\label{stringendingonwall}

In this  section we consider an Abelian string ending on the
elementary wall. The $12\to 14$ wall  has a nonvanishing
$r$-component of the first flavor
inside the wall, see Sect. \ref{elwall}. Therefore, only the
(0,1)-string
whose flux is orthogonal to the $r$-weight vector can end on
this wall.

Needless to say  the solution of the first-order equations (\ref{foej})
for the string
ending on the wall can be found only numerically, especially near the
end-point of
the string  where both the string and the wall profiles are heavily deformed.
However, far   from the string end-point, deformations are
weak and we can find the asymptotic behavior analytically.

Let the string be on the $z<0$ side of the wall, inside the 12-vacuum.
Consider first the region $z\to -\infty$ far away
from the string end-point at  $z\sim 0$. Then the solution to
(\ref{foej}) is given by an almost unperturbed string.
Namely, at $z\to-\infty$ there is no $z$ dependence
to the leading order, and, hence,    the solution
\beqn
\vp^{kA}
&=&
\left(
\begin{array}{cc}
\phi_2(r) & 0  \\[2mm]
0 &  \phi_1(r) \\
\end{array}\right) ,
\nonumber\\[3mm]
A^3_{i}(x)
&=&
-\epsilon_{ij}\,\frac{x_j}{r^2}\,
f_3(r)\, ,
\nonumber\\[3mm]
A_{i}(x)
&=&
\epsilon_{ij}\,\frac{x_j}{r^2}\,
f(r),
\label{s01}
\eeqn
(which is a singular-gauge version of the solution (\ref{sol})
for $n=0$, $k=1$, cf. Eq.~(\ref{s10})) satisfies Eqs.~(\ref{foej}).
We also take the fields  $A_{3}=A_{3}^a=0$  and  $\vp^B$ ($B=3,4$)
to be zero, with  $a$'s equal  to  their VEV's (\ref{avev}).
On the other side of the wall, at $z\to +\infty$,
we have an almost unperturbed 14-vacuum with the fields  given by their
respective VEV's.

Now consider the domain $r\to\infty$ at small $z$. In this domain the solution
to Eq.~(\ref{foej}) is given by a  perturbation of the wall solution. Let us
use the {\em ansatz} in which the solutions for the fields $a$, $a^a$ and
 $\vp^{A}$ are
 given by
the same equations (\ref{a}), (\ref{r0}), (\ref{wq2}) and (\ref{wqB})
 in which the
size of the wall is still given by  (\ref{R}), and
 {\em the only modification} is
that the position of the wall $z_0$ and the phase $\sigma$  now become
slowly-varying functions of $r$ and $\alpha$ (i.e. the polar coordinates
on the $(x_1,x_2)$ plane). It is quite obvious that $z_0$ will depend
only on $r$.

As long as  the third,   fourth and   fifth  equations in
 (\ref{foej}) do not contain
derivatives with respect to $x_i$, $i=1,2$, they are identically satisfied
 for any functions $z_0(r,\alpha)$ and $\sigma (r,\alpha)$ (note, that
$F^{*}_3=F^{*a}_3 =0$,
the field strength is almost parallel to  the  domain wall plane
and  $A_3= A_3^a=0$).

However, the first  two  and the last two equations in (\ref{foej})
 become nontrivial.
Consider the first two. Inside the string the gauge fields are
directed along the $z$ axis and their fluxes are   $2\pi$ for
$F_3^{*}$ and $-2\pi$ for $F_3^{*3}$ (remember, we treat the (0,1)-string).
 This flux is spread out  inside the wall and
directed  almost along $x_i$ in the $(x_1,x_2)$ plane at large $r$.
Since the flux  is conserved,  we have
\beqn
F^{*}_{i}
&=&
\frac{1}{R}\,\, \frac{x_{i}}{r^2}\,
\nonumber\\[3mm]
F^{*3}_{i}
&=&
-\frac{1}{R}\,\, \frac{x_{i}}{r^2}\,,
\label{flux}
\eeqn
inside the wall at $|z-z_0(r,\alpha)|<R/2$.

Substituting this in the first two equations in (\ref{foej}) and assuming
that $z_0$ depends only on $r$ we  get that the two equations are
consistent with each other and
\beq
\label{derz}
\pt_{r}z_0=\frac1{\Delta m r}\, .
\eeq
Needless to say   our ``adiabatic" approximation holds
only provided the above derivative is small, i.e.
sufficiently far from the string end point,
 $\sqrt{\xi} r\gg1$.

The solution to this equation is straightforward,
\beq
\label{z0}
z_0=\frac1{\Delta m }\ln {r} + {\rm const}.
\eeq
We see that the wall is logarithmically bent according to the Coulomb law
in 2+1 dimensions.
Similar to the case considered in \cite{Shifman:2002jm}, one can show that
 this bending produces a balance of forces
between the string and the wall in the $z$ direction so that the whole
configuration is
static.

Now let us consider the last equation in (\ref{foej}). First, we will dwell
on  the gauge
potential which enters the covariant derivatives in this equation. In order to
produce the field strength (\ref{flux}), $A_\mu$ and   $A^a_\mu$
in the middle domain should
reduce to
\beqn
A_{i} &=& \frac{1}{R}\,\ve_{ij}\frac{x_{j}}{r^2}\left[
z-z_0(r)+\frac{R}{2}\right]
\; , \qquad i=1,2\, ,
\nonumber\\[4mm]
A_{i}^3 &=& -\frac{1}{R}\,\ve_{ij}\frac{x_{j}}{r^2}\left[
z-z_0(r)+\frac{R}{2}\right]
\,, \qquad i=1,2\, ,
\nonumber\\[4mm]
 A_{0} &=& A_{0}^3=0,\qquad A_3=A_{3}^3=0\, .
\label{gaugepot}
\eeqn
Please,  observe that nonvanishing field of the first quark flavor satisfies the
equation since it has only the
$r$-component which is not charged with respect to  the field (\ref{gaugepot}).
Consider the second quark flavor whose $b$-component is given by
(\ref{wq2}) in  the middle domain
near the left edge of the wall at $z-z_0 \sim -R/2$. Taking into account
the gauge potentials (\ref{gaugepot}) and the wall bending
(\ref{derz}) it is easy to check that
the second flavor satisfies the last equation in (\ref{foej}).

Finally, let us consider $\vp^{b4}$  in the middle domain
near the right edge of the wall, $z-z_0 \sim R/2$. Substituting
(\ref{wqB}) into the last equation in (\ref{foej}) we get the following
equations for the phase $\sigma$:
\beq
\label{dersig}
\frac{\pt \sigma}{\pt \alpha}=1,\qquad \frac{\pt \sigma}{\pt r}=0\, .
\eeq
The solution to these equations is
\beq
\label{sigmaalpha}
\sigma =\alpha\, .
\eeq
 In terms of
the dual  Abelian gauge field localized on the wall, this solution reflects
nothing but the unit source charge.

The above relation between the vortex solution and the unit source charge
calls for a comment. One can identify the
compact scalar field $\sigma$ with the electric  field living
on the domain wall world volume via (\ref{21gaugenorm}). Then
the result (\ref{sigmaalpha}) gives  for this electric
field
\beq
\label{elfnormp}
F_{0i}^{(2+1)}=\frac{e^2_{2+1}}{2\pi}
\,\,\frac{x_i}{r^2}\,,
\eeq
where the (2+1)-dimensional coupling is given in Eq.~(\ref{21coupling}).

This  is the field of a point-like electric charge
in 2+1 dimensions placed at $x_i=0$.
The interpretation of this result is that the string end-point
 on the wall plays a role of the electric charge in
the  dual U(1) theory on the wall, cf. \cite{Shifman:2002jm}.

\subsection{Non-Abelian string ending on the composite wall}
\label{naseotcw}

Now we pass to the non-Abelian string ending on the composite wall
interpolating between the 12- and 34-vacua. Our strategy is as follows.
We start with the (0,1) Abelian string as in Sect.~\ref{stringendingonwall}
and consider its junction with the composite $12\to 34$ wall of Sect.~\ref{1234}.
We then  apply the SU(2)$_{C+F}$ rotation introduced and discussed
in Sect.~\ref{naftin2} to this junction.
Namely, we write down the first two flavors as a $2\times 2$ matrix
$\vp^{kA}$ ($A=1,2$) and the last two flavors as a $2\times 2$ matrix
$\vp^{kB}$ ($B=3,4$), and
 rotate them according to
\beqn
\vp
&\to&
 U_{C+F}\;\vp\;U_{C+F}^{-1}, \; {\rm flavor\;indices}=1,2\,,
\nonumber\\[3mm]
\vp
&\to&
U_{C+F}\;\vp \;U_{C+F}^{-1}, \; {\rm flavor\;indices}=3,4\,,
\label{c+f}
\eeqn
with one and the same matrix from SU(2)$_{C+F}$.

Note that both the 12- and 34-vacua do  not break this symmetry.
However,  the string  and the string-wall junction are not
invariant.
Therefore, if we apply this rotation to the solution for the
(0,1)-string ending
on the composite wall we will get the solution of (\ref{foej}) for a
non-Abelian
string ending on the composite wall, with the same energy.

\begin{figure}
\epsfxsize=5cm
\centerline{\epsfbox{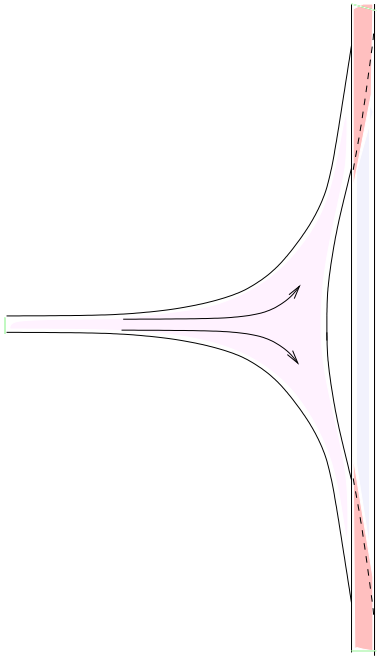}}
\caption{ Junction of the string and the  composite wall. The string
pulls out one of the components of the composite wall. Arrows show
the spread of the magnetic flux inside the wall.    }
\label{fi:junction}
\end{figure}

Having set the general strategy,
it is time to proceed to a technical
analysis of the junction of the (0,1) string with the composite wall.
Assume for simplicity that in the absence of the string, the matrix
$\tilde{U}=I$ (see Eq.~(\ref{wq34}))
so  that the deviation of $\tilde{U}$ from the unit matrix is
due to the string flux.
Our composite wall can be considered as
a marginally stable bound state of the $12\to 14$
and $14\to 34$ walls.
While  the solution presented in
Sect.~\ref{1234} has a vanishing separation
between the constituents,
the two elementary walls
in this bound state do not interact and their positions
can be shifted to arbitrary separations. As  the (0,1)-string can end
only on the $12\to 14$ wall, it is clear that it will pick up this wall
and pull it out to the left, see Fig.~\ref{fi:junction}.
The $(14\to 34)$  constituent stays
unbent and does not play a  role in the junction solution at
hand.\,\footnote{Of course, there could be some interaction of the end point
of the string on the $12\to 1B$ wall with  $14\to 34$ wall but this
interaction is short range and dies out at $r\gg 1/\sqrt{\xi}$.}
We see that
the solution for the (0,1)-string ending on the composite wall reduces to
the solution for the (0,1)-string ending on the $(12\to 14)$ wall considered in
Sect.~\ref{stringendingonwall}.

The solution for the (0,1)-string ending on the $12\to 14$ wall has
the $b$-component of the fourth flavor multiplied by $\exp{(-i\sigma)}$
with $\sigma$ given by (\ref{sigmaalpha}), see Eq.~(\ref{wqB}). The
$14\to 34$ wall has all phases vanishing because there is no flux going
inside this wall.

Thus, our junction
has the quark matrix of the final vacuum determined  by the matrix
\beq
\label{ut}
\tilde{U}=\left(
\begin{array}{cc}
1 & 0\\[2mm]
0 & e^{-i\alpha}\\
\end{array} \right),
\eeq
see Eq.~(\ref{wq34}).
This shows that the junction of the (0,1)-string and
the composite wall  has
the following  phase
$\sigma_0$ and the SU(2) matrix $U$ (see (\ref{flmat})):
\begin{eqnarray}
\sigma^0
&=&
-\frac{\alpha}{2}\, ,
\nonumber\\[3mm]
U
&=&
\exp{\left(i\, \frac{\tau^3}{2}\alpha\right)}\, .
\label{sigmascom}
\end{eqnarray}

Now let us apply the SU(2)$_{C+F}$ rotation (\ref{c+f})
to the whole configuration. The flux of the string is now
determined by an arbitrary vector $n^a$ inside the SU(2) subgroup
while the quark matrix of the final vacuum gets rotated as
\beq
\label{utrot}
\tilde{U}=U_{C+F}\left(
\begin{array}{cc}
1 & 0\\[2mm]
0 & e^{-i\alpha}
\end{array} \right)
U_{C+F}^{-1}\, .
\eeq
In terms of the phase $\sigma_0$ and the SU(2) matrix $U$ this amounts to
\beqn
\sigma^0
&=&
-\frac{\alpha}{2}\, ,
\nonumber\\[3mm]
U
&=&
\exp{\left( i\, \frac{n^a\tau^a}{2}\alpha\right)}\, .
\label{sigmascomp}
\eeqn

This result clearly means that the end-point of the string is a point-like
source of the  non-Abelian gauge field on the composite wall.
To see this more explicitly let us write down the (2+1)-dimensional
gauge fields associated with  $\sigma_0$ and the  matrix $U$ using
Eq.~(\ref{21gaugefields}). We get
\beqn
F_{0i}^{(2+1)}
&=&
-\frac{e^2_{2+1}}{4\pi}
\,\,\frac{x_i}{r^2}\,,
\nonumber\\[4mm]
F_{0i}^{a(2+1)}
&=&
\frac{g^2_{2+1}}{2\pi}
\,\,\frac{x_i}{r^2}\,n^a\, .
\label{21charge}
\eeqn
As we see, this is the field of a classical point-like charge
in the SU(2)$\times$U(1) gauge theory on the wall. The
direction of the SU(2) field
in the color space is determined by the vector $n^a$ associated
with the string flux.

\subsection{Gauge symmetry breaking}
\label{GSB}

As was discussed at the end of Sect.~\ref{cowo},
if  our composite
wall is split into elementary components whose separation is
larger than their thickness, the non-Abelian gauge
symmetry in the world-volume theory  (\ref{21theoryna})
is broken down to  U(1)$\times$U(1). In particular, the mass of
the (2+1)-dimensional $W$-bosons becomes proportional to
the separation $l$ between the elementary walls,
see Eq.~(\ref{21Wmass}).

Our analysis demonstrates that  localization of a gauge field on a wall
and existence of the corresponding
string-wall junction are two sides of one and the same phenomenon.
In this section we address the question: ``what happens with the
string-wall junction
in the (3+1)-dimensional bulk theory  if we split
the composite wall and pull the components apart?"

Consider a string-wall junction for the
non-Abelian string ending on the composite wall, as in Sect.~\ref{naseotcw}.
If $\langle l\rangle =0$ two elementary walls which
form the $12\to 34$ wall overlap at large $r$
($r$ is the distance from the string end-point along the wall). In fact,
the composite $12\to 34$ wall  can be represented as a bound state of two
elementary walls in many different ways depending on which particular
combination of the quark fields is non-zero in the given
elementary walls. In particular,
the string with flux $\sim n^a$ picks up a particular elementary
wall with
\beq
\label{nzq}
\vp^{kA}\sim U_{C+F}\left(
\begin{array}{cc}
1 & 0\\
0 & 0\\
\end{array} \right)U_{C+F}^{-1},\;\; A=1,2\,,
\label{mrmr}
\eeq
non-vanishing inside the wall. The string pulls it out to
the left while the ``other wall" is  (almost) unbent, see Fig.~\ref{fi:junction}.
The string ends on the wall specified by
(\ref{mrmr}) so the string flux spreads
inside this wall, and at large $r$ is given by (\ref{21charge}). The flux
direction in the color space is determined by the string flux
vector $n^a$.

\begin{figure}[h]
\epsfxsize=5cm
\centerline{\epsfbox{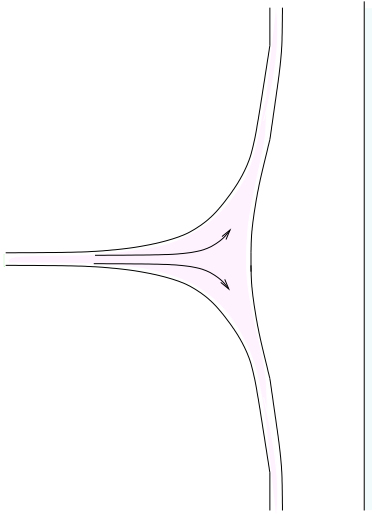}}
\caption{Junction of the string and the  composite wall
for $\langle l\rangle\ne 0$.     }
\label{fi:junctionbr}
\end{figure}

Now suppose that $\langle l\rangle \ne 0$. In other words,
the composite wall is split in particular elementary
components which do not overlap even at $r\to\infty$, see
Fig.~\ref{fi:junctionbr}.
Say, if we have the 14-vacuum between the walls
the  elementary wall on the left has a concrete non-vanishing quark
field, with necessity, namely, the quark field proportional to
\beq
\label{nzql1}
\vp^{kA}\sim \left(
\begin{array}{cc}
1 & 0\\
0 & 0\\
\end{array} \right),\;\; A=1,2.
\eeq
This guarantees that only the (0,1)-string
can end on the
wall. The flux inside the wall is given in this case by
Eq.~(\ref{21charge})
with a specific $n^a$, namely,  $n^a=(0,0, 1)$.

If, instead, we have the 23-vacuum between the
walls the elementary wall on the left
has the non-vanishing quark field proportional to
\beq
\label{nzql2}
\vp^{kA}\sim \left(
\begin{array}{cc}
0 & 0\\
0 & 1\\
\end{array} \right),\;\; A=1,2.
\eeq
This means  that only the (1,0)-string
can end on this wall configuration.
The flux inside the wall will be given in this case by
Eq.~(\ref{21charge}) with  $n^a=(0,0,-1)$.
String with generic fluxes
determined by an
arbitrarily-oriented vector $n^a$ just cannot end on the composite
domain wall if $\langle l\rangle > R $.

Of course, this is perfectly consistent with the breaking of
the SU(2)$\times$U(1) gauge
group down to U(1)$\times$U(1),
so that the (2+1)-dimensional $W^\pm$-bosons are
heavy, do not propagate,
and the massless gauge fields on the wall are
by $A_{n}^{2+1}$ and $A_{n}^{3(2+1)}$.

\section{Dynamics of the world-volume theory}
\label{dotwvt}
\setcounter{equation}{0}

In this section we briefly discuss   dynamics of  the world-volume
theory emerging on the wall. We will focus on non-perturbative
instanton effects which lead to a run-away vacuum
in the world-volume theory.

\subsection{BPS saturation of the composite wall}
\label{bpsscw}

In Sect.~\ref{foe} we demonstrated that the
central charges of the composite wall and its two
constituents are aligned, so that the tension
of the composite wall equals twice the tension of the elementary wall.
This statement is valid to any order in perturbation theory.
If so, the composite wall would be marginally stable:
there would be no interaction between the constituent walls
no matter what the separation between the constituents is.

In terms of the world-volume theory (\ref{21theoryna})
this means that the flat direction is not lifted.
No superpotential is generated to any finite order in
perturbation theory. Phrased this way, the assertion seem
almost obvious. From other examples we know, however, that
a superpotential might be generated non-perturbatively.
An indication that this may be the case comes from
the occurrence of the anomalous terms in the central charge (\ref{stenkaz}). Some well-known old results
will allow us to answer this question quickly.

\subsection{Non-perturbative effects}
\label{npeff}

Dynamics of \ntwo (2+1)-dimensional gauge theory with the SU(2)
gauge group was studied by Affleck, Harvey and Witten \cite{AHW}. It was
shown that instantons (in (2+1) dimensions
they are nothing but 't Hooft-Polyakov monopoles, tHP for short)
generate a superpotential which produces a run-away
vacuum.\footnote{Historically this work presented the first example
ever in which   perturbative non-renor\-malization theorem --
the absence of the superpotential -- was shown to be violated
non-per\-turbatively.}
Classically there is a flat direction in the theory (\ref{21theoryna})
so that the scalar field $a_{2+1}^3$ can develop an arbitrary VEV breaking
the SU(2) gauge group on the wall
down to U(1). Then $A_{n}^{1,2(2+1)}$ acquire a
mass while $A_{n}^{2+1}$ and $A_{n}^{3(2+1)}$ remain massless.
We can dualize $A_{n}^{3(2+1)}$ into a compact scalar $\sigma^3$
according to (\ref{21gaugefields}), to introduce a complex scalar filed
$\Phi =a_{2+1}^3+i\sigma^3$. This scalar is the  lowest component of a
chiral supermultiplet.

It was shown in \cite{AHW} that instantons-tHP monopoles generate the
superpotential\,\footnote{The mechanism is quite similar to the one in the
non-supersymmetric version of the theory studied by Polyakov
 \cite{Polyakov:1976fu}.
Monopoles form a Coulomb gas in (2+1) dimensions which is equivalent
to the sine-Gordon theory.}
\beq
\label{21sup}
W_{2+1}\sim \exp{\left(-\frac{\Phi}{g^2_{2+1}}\right)}.
\eeq
The potential arising from this superpotential
\beq
\label{21pot}
V_{2+1}\sim \exp{\left(-\frac{2a_{2+1}^3}{g^2_{2+1}}\right)}
\eeq
leads to a run-away vacuum.
Using (\ref{separationid}) we can reinterpret this potential
as a interaction potential between elementary walls which
comprise our composite $(12\to 34) $ wall,
\beq
\label{wallinter}
V_{int}\sim \exp{\left(-\frac{g_2}{g_1}\frac{\Delta m}{\pi}\;l\right)}.
\eeq

 Classically the elementary walls do not interact.
Non-perturbative effects on the world volume induce a repulsive interaction
between the elementary walls, so that the BPS bound state can formally appear
only in the limit of infinite separation between walls.
However, in fact, the
interactions (\ref{wallinter}) become negligibly small
 already at
separations $l_*$ of the order of
$(\Delta m)^{-1}$. In other words, the ratio $l_*/R\sim \xi /(\Delta m)^{2}\ll1$.
Here $R$ represents the wall thickness.

We observe  an interesting interplay between bulk physics
and physics on the wall. In particular, above we extracted the interaction
potential between the elementary walls from the known results
on  the effective theory on the wall. This bulk/brane duality
is somewhat similar to the AdS/CFT correspondence. A weak
coupling regime in the bulk maps onto a strong coupling regime on the wall
and {\em vice  versa}. To see that this is indeed the case,
suffice it to remember that when the bulk coupling constants
$g^2_1$ and $g^2_2$ are small the (2+1)-dimensional couplings
 (\ref{21couplings}) are large   compared to the characteristic scale
of massive excitations on the wall which are of the order of $ 1/\tilde{R}$
(cf. \cite{Shifman:2002jm}).

\subsection{Compatibility with the D-brane picture}
\label{cwtdbp}

Returning to the issue of the elementary wall exponential repulsion,
one may ask how this  can be interpreted in view of the well-known fact
that   the two-stacks (as well as all other stacks) of  D-branes
are stable. The answer is quite clear. D-branes have no thickness.
Our construction belongs to weak coupling
where the walls do have a thickness. The repulsive non-perturbative interaction
dies off at distances much less than the wall thickness.
Therefore, squeezing the walls to vanishing thickness
automatically switches the repulsion off.

\section{Conclusions}
\label{conclu}

In this paper we studied   localization of  non-Abelian gauge fields
on domain walls. We showed that although elementary domain walls can localize
only Abelian fields the composite domain wall does localize non-Abelian
gauge fields. In order to have this localization we considered \ntwo
QCD with the gauge group SU(2)$\times$U(1) in a   special regime.   Although
the gauge group is completely Higgsed by the quark VEV's in the bulk it is restored
inside the composite domain wall where all quark fields are almost zero.
This ensures   localization of the non-Abelian gauge field on the wall.

Another side of this phenomenon is the possibility for  non-Abelian
flux tubes to end on the wall. The non-Abelian flux tubes were recently
found in Ref. \cite{Auzzi:2003fs} in four dimensions
and in Ref. \cite{Hanany:2003hp} in three dimensions. They carry
additional orientational zero modes corresponding to rotations
of the  color-magnetic flux inside the SU(2) subgroup of the gauge group.
The key ingredient for the existence of such non-Abelian strings is
the presence of a diagonal color-flavor group SU(2)$_{C+F}$ unbroken by
the vacuum condensates (color-flavor locking). We found a 1/4-BPS solution
for such  non-Abelian string ending on the composite wall. The end point
of the string plays the role of a color charge in the (2+1)-dimensional
(dual) non-Abelian gauge theory on the wall.

To study the string-wall junctions we use the first-order master equations
(\ref{foej}) which in the Abelian case were
derived in \cite{Shifman:2002jm}. In fact, the same
equations can be used  for  all possible junctions between domain
walls, strings and monopoles. In particular, recently they were
used \cite{Tong:2003pz} to study
the $(1,-1)$ monopole as a junction of the (1,0) and (0,1) strings
 in the limit of large $\Delta m$. We discuss this
monopole in the
opposite limit of equal quark masses, $\Delta m\to 0$,  when it becomes a
junction of two strings associated with two quantum vacua of the
(1+1)-dimensional O(3)
sigma model on the string world sheet.
 We show that the mass of this monopole is lifted from zero by
non-perturbative effects in the O(3) sigma model. We will
come back to this issue  \cite{SYfut}.

We also  studied the effective (2+1)-dimensional non-Abelian theory on
 the composite domain wall. We found an interesting
bulk/brane duality. In particular, the weak coupling regime in the bulk
maps onto the strong coupling regime on the wall and {\em vice versa}. This is
quite similar in spirit to the Ads/CFT correspondence.

\section*{Acknowledgments}

We are grateful to A. Gorsky,  Adam Ritz and
Arkady Vainshtein for  discussions.
The work of M.~S. is supported  by DOE grant DE-FG02-94ER408, A.~Y.
is supported in part by the Russian Foundation for Basic
Research   grant No.~02-02-17115, by INTAS grant
No.~00-00334 and by Theoretical Physics Institute
at the University of Minnesota.


\newpage

\begin{thebibliography}{99}
\addcontentsline{toc}{section}{References}

\bibitem{P}
J.~Polchinski,
Phys.\ Rev.\ Lett.\  {\bf 75}, 4724 (1995)
[hep-th/9510017];
see also the excellent text by J.~Polchinski, {\em String Theory}, Vols. 1 and 2
(Cambridge University Press,  Cambridge, 1998).

\bibitem{DS}
G.~R.~Dvali and M.~A.~Shifman,
Phys.\ Lett.\ B {\bf 396}, 64 (1997)
(E)\ B {\bf 407}, 452 (1997)
[hep-th/9612128].

\bibitem{Witten:1997ep}
E.~Witten,
Nucl.\ Phys.\ B {\bf 507}, 658 (1997)
[hep-th/9706109].

\bibitem{Shifman:2002jm}
M.~Shifman and A.~Yung,
Phys.\ Rev.\ D {\bf 67}, 125007 (2003)
[hep-th/0212293].

\bibitem{Hanany:2003hp}
A.~Hanany and D.~Tong,
JHEP {\bf 0307}, 037 (2003)
[hep-th/0306150].

\bibitem{Auzzi:2003fs}
R.~Auzzi, S.~Bolognesi, J.~Evslin, K.~Konishi and A.~Yung,
Nucl. Phys. {\bf B673}, 187  (2003)
hep-th/0307287.

\bibitem{Tong:2003pz}
D.~Tong,
{\em Monopoles in the Higgs phase,}
hep-th/0307302.

\bibitem{SW1}
N.~Seiberg and E.~Witten, Nucl. Phys. {\bf B426}, 19 (1994) [hep-th/9407087].

\bibitem{SW2}
N.~Seiberg and E.~Witten, Nucl. Phys. {\bf B431}, 484  (1994)
[hep-th/9408099].

\bibitem{DV}
G.~Dvali and A.~Vilenkin,
Phys.\ Rev.\ D {\bf 67}, 046002 (2003)
[hep-th/0209217].

\bibitem{APS}
P.~Argyres, M.~Plesser and N.~Seiberg, Nucl. Phys. {\bf B471}, 159  (1996)
[hep-th/9603042].

\bibitem{CKM}
G.~Carlino, K.~Konishi and H.~Murayama,
Nucl. Phys. {\bf B590}, 137  (2000) [hep-th/0005076].

\bibitem{MY}
A.~Marshakov and A.~Yung,
Nucl.\ Phys.\ B {\bf 647}, 3 (2002)
[hep-th/0202172].

\bibitem{ABEK}
R.~Auzzi, S.~Bolognesi, J.~Evslin and  K.~Konishi,
{\em Nonabelian monopoles and vortices that confine them,}
hep-th/0312233.

\bibitem{HSZ}
A.~Hanany, M.~Strassler and A.~Zaffaroni,
Nucl. Phys. {\bf B513},  87   (1998) [hep-th/9707244].

\bibitem{VY}
A.~I.~Vainshtein and A.~Yung,
Nucl.\ Phys.\ B {\bf 614}, 3 (2001)
[hep-th/0012250].

\bibitem{EFMG}
J.~D.~Edelstein, W.~G.~Fuertes, J.~Mas and J.~M.~Guilarte,
Phys.\ Rev.\ D {\bf 62}, 065008 (2000)
[hep-th/0001184].

\bibitem{ANO}
A.~Abrikosov, Sov.~Phys. JETP {\bf32} 1442  (1957)
[Reprinted in {\em Solitons and Particles}, Eds. C. Rebbi and G. Soliani
(World Scientific, Singapore, 1984), p. 356];
H.~Nielsen and P.~Olesen, Nucl.~Phys. {\bf B61} 45 (1973)
[Reprinted in {\em Solitons and Particles}, Eds. C. Rebbi and G. Soliani
(World Scientific, Singapore, 1984), p. 365].

\bibitem{deAzcarraga:gm}
J.~A.~de Azcarraga, J.~P.~Gauntlett, J.~M.~Izquierdo and P.~K.~Townsend,
Phys.\ Rev.\ Lett.\  {\bf 63}, 2443 (1989).

\bibitem{KSS}
A.~Kovner, M.~A.~Shifman and A.~Smilga,
Phys.\ Rev.\ D {\bf 56}, 7978 (1997)
[hep-th/9706089].

\bibitem{Chibisov:1997rc}
B.~Chibisov and M.~A.~Shifman,
Phys.\ Rev.\ D {\bf 56}, 7990 (1997);
(E) \ D {\bf 58}, 109901 (1998);
[hep-th/9706141].

\bibitem{Ritz:2002fm}
A.~Ritz, M.~Shifman and A.~Vainshtein,
Phys.\ Rev.\ D {\bf 66}, 065015 (2002)
[hep-th/0205083].

\bibitem{Shifman:1999mv}
M.~A.~Shifman and A.~I.~Vainshtein,
{\em Instantons versus supersymmetry: Fifteen years later},
in M. Shifman, {\em ITEP Lectures on Particle Phsyics and Field Theory},
(World Scientific, Singapore, 1999), Vol. 2, p. 485
[hep-th/9902018].

\bibitem{GS}
A.~Gorsky and M.~A.~Shifman,
Phys.\ Rev.\ D {\bf 61}, 085001 (2000)
[hep-th/9909015].

\bibitem{Gabadadze:1999pp}
G.~Gabadadze and M.~A.~Shifman,
Phys.\ Rev.\ D {\bf 61}, 075014 (2000)
[hep-th/9910050].

\bibitem{at}
E.~R.~Abraham and P.~K.~Townsend,
Nucl.\ Phys.\ B {\bf 351}, 313 (1991).

\bibitem{HS}
Z.~Hlousek and D.~Spector,
Nucl.\ Phys.\ B {\bf 370}, 143 (1992);
J.~Edelstein, C.~Nu\~{n}ez and F.~Schaposnik,
Phys.\ Lett.\ B {\bf 329}, 39 (1994)
[hep-th/9311055].

\bibitem{DDT}
S.~C.~Davis, A.~C.~Davis and M.~Trodden,
Phys.\ Lett.\ B {\bf 405}, 257 (1997)
[hep-ph/9702360].

\bibitem{FI}
P.~Fayet and J.~Iliopoulos,
Phys.\ Lett.\ B {\bf 51}, 461 (1974).

\bibitem{FP}
S.~Ferrara and M.~Porrati,
Phys.\ Lett.\ B {\bf 423}, 255 (1998)
[hep-th/9711116].

\bibitem{Haag:1974qh}
R.~Haag, J.~T.~Lopuszanski and M.~Sohnius,
Nucl.\ Phys.\ B {\bf 88}, 257 (1975).

\bibitem{SYfut}
M.~Shifman and A.~Yung,
{\em Non-Abelian String Junctions as Confined Monopoles,}
hep-th/0403149.

\bibitem{Losev:2003gs}
A.~Losev and M.~Shifman,
Phys.\ Rev.\ D {\bf 68}, 045006 (2003)
[hep-th/0304003].

\bibitem{B}
E.~B.~Bogomolny,
Yad.\ Fiz.\  {\bf 24}, 861 (1976) [Sov.\ J.\ Nucl.\ Phys.\  {\bf 24}, 449 (1976),
reprinted in {\em Solitons and Particles}, Eds. C. Rebbi and G. Soliani
(World Scientific, Singapore, 1984), p. 389].

\bibitem{OW}
E.~Witten and D.~I.~Olive,
Phys.\ Lett.\ B {\bf 78}, 97 (1978).

\bibitem{Polyakov:1976fu}
A.~M.~Polyakov,
Nucl.\ Phys.\ B {\bf 120}, 429 (1977).

\bibitem{Shifman:2002yi}
M.~Shifman and A.~Yung,
Phys.\ Rev.\ D {\bf 66}, 045012 (2002)
[hep-th/0205025].

\bibitem{VS}
H.J.~de Vega and F.A.~Shaposnik, Phys. Rev. Lett. {\bf 56},  2564 (1986);
Phys. Rev. {\bf D34},  3206  (1986).

\bibitem{HV}
J.~Heo and T.~Vachaspati, Phys. Rev. {\bf D58},   065011 (1998)
[hep-th/9801455].

\bibitem{Su}
P.~Suranyi,
Phys.\ Lett.\ B {\bf 481}, 136 (2000)
[hep-lat/9912023].

\bibitem{SS}
F.A.~Shaposnik and P.~Suranyi, Phys. Rev. {\bf D62},  125002 (2000)
[hep-th/0005109].

\bibitem{KB}
 M.~Kneipp and P.~Brockill,
Phys. Rev. {\bf D64},   125012 (2001)  [hep-th/0104171].

\bibitem{KS}
K.~Konishi and L.~Spanu,
Int.\ J.\ Mod.\ Phys.\ A {\bf 18}, 249 (2003)
[hep-th/0106175].


\bibitem{BarH}
K.~Bardakci and M.~B.~Halpern,
Phys.\ Rev.\ D {\bf 6}, 696 (1972).

\bibitem{Alvarez}
L.~Alvarez-Gaum\'{e} and D.~Z.~Freedman,
Commun.\ Math.\ Phys.\  {\bf 91}, 87 (1983).

\bibitem{Gates:1983py}
S.~J.~Gates,
Nucl.\ Phys.\ B {\bf 238}, 349 (1984).

\bibitem{Gates:nk}
S.~J.~Gates, C.~M.~Hull and M.~Ro\v{c}ek,
Nucl.\ Phys.\ B {\bf 248}, 157 (1984).

\bibitem{HoVa}
P.~Fendley and K.~A.~Intriligator,
Nucl.\ Phys.\ B {\bf 380}, 265 (1992)
[hep-th/9202011];
S.~Cecotti and C.~Vafa,
Commun.\ Math.\ Phys.\  {\bf 158}, 569 (1993)
[hep-th/9211097];
K.~Hori and C.~Vafa,
{\em Mirror symmetry,}
hep-th/0002222.

\bibitem{We}
E.~Weinberg,
Nucl.\ Phys.\ B {\bf 167}, 500 (1980);
Nucl.\ Phys.\ B {\bf 203}, 445 (1982).


\bibitem{AHW}
I.~Affleck, J.~A.~Harvey and E.~Witten,
Nucl.\ Phys.\ B {\bf 206}, 413 (1982).


\end{thebibliography}
\end{document}